\documentclass[3p,authoryear]{elsarticle}

\usepackage{amsmath,amsthm,amsfonts,amssymb}

\usepackage{XCharter}
\usepackage[charter,vvarbb]{newtxmath}
\usepackage{graphicx}
\usepackage{booktabs}
\usepackage{enumitem}
\usepackage{rotating}
\usepackage{wrapfig}

\newsavebox{\tabbox}

\newlength{\tabw}
\usepackage{tikz}

\usepackage[section,below]{placeins}
\extrafloats{64}

\usepackage{caption}
\usetikzlibrary{arrows.meta,positioning,fit,backgrounds,calc}
\usepackage{url}
\usepackage[hidelinks]{hyperref}
\hypersetup{pdfauthor={},pdftitle={},pdfsubject={},pdfkeywords={},pdfcreator={}}

\newtheorem{theorem}{Theorem}
\newtheorem{proposition}[theorem]{Proposition}

\newtheorem{remark}{Remark}

\newcommand{\titledParagraph}[1]{\noindent\textbf{#1}}

\newcommand{\tabref}[1]{\textbf{Table~\ref{#1}}}
\newcommand{\figref}[1]{\textbf{Fig.~\ref{#1}}}
\newcommand{\figrefp}[2]{\textbf{Fig.~\ref{#1}(#2)}}

\journal{Transportation Research Part E}

\begin{document}

\begin{frontmatter}

\title{Marginal-Cost Clustering for Courier Dispatch in Pickup and Delivery
Problems with Time Windows and Fleet Sizing}

\author[njit]{SangWoo Park}
\ead{sangwoo.park@njit.edu}
\author[plt]{Elson Cibaku}
\ead{eciba001@plattsburgh.edu}
\author[njit]{Sanchoy Das}
\ead{sanchoy.das@njit.edu}
\affiliation[njit]{organization={Department of Mechanical and Industrial
                   Engineering, New Jersey Institute of Technology},
                   city={Newark}, state={NJ}, country={USA}}
\affiliation[plt]{organization={Department of Management, Information Systems
                   \& Analytics, State University of New York at Plattsburgh},
                   city={Plattsburgh}, state={NY}, country={USA}}
                   
\begin{abstract}
Courier dispatch poses a pickup and delivery problem with time windows in which
requests are batched, windows are soft, couriers start where they are and take
open routes, and a fixed dispatch charge makes fleet size a decision. We contribute to the cluster-first, route-second family a
grouping criterion that scores a request by its \emph{marginal cost}, the
travel and lateness it adds to the route a courier already plans, rather than by
its distance to a centroid, one point standing for a whole workload in space
and time. The marginal cost instead knows where a courier is, when it
becomes free and what it already carries. Each pass, we hold the marginal costs fixed while we solve
one integer program that assigns every request to a courier and chooses which to
open, then improve its solution by a descent that relocates and swaps requests
between couriers. We then route each dispatched courier exactly.
That subproblem is naturally a mixed-integer program, and we solve it by label
setting over visited sets, which we prove optimal and cheap enough to run every
pass. On instances we generate and on adapted Li \& Lim
benchmark batches, we beat the best of six published clustering rules by
29\% as published and by 12\% once those rules carry the same descent. Against an adaptive large neighbourhood search at matched wall-clock we
are ahead on the generated instances and level on the benchmark, and we handle
1000 requests and 300 couriers in about a minute.
\end{abstract}


\end{frontmatter}

\section{Introduction}\label{sec:intro}

\noindent On-demand delivery platforms dispatch batches of orders to a pool of
couriers every few minutes. Each order is a pickup at a merchant and a delivery
to a customer, each has a promised time window, each courier can hold only a few
orders at once, and putting one more courier on the road costs money. The
dispatcher must decide, in the time between batches, which courier takes which
orders, in what sequence, and how many couriers to use at all.

A monolithic mixed-integer formulation of this problem conflates two different
difficulties. \emph{Assignment}, which courier serves which request, forces
a routing variable for every (stop, stop, courier) triple, giving
$\Theta(|\mathcal{R}|^2|\mathcal{K}|)$ binaries. \emph{Sequencing}, the order
in which one courier visits its own stops, is a travelling-salesman problem
with precedence and time windows and cannot be reformulated away. Fixing the
assignment removes the first exactly and shrinks the second to
$|\mathcal{K}|$ independent problems on a few stops each, small enough to solve
to optimality. That is the cluster-first, route-second idea, and it is old.

What is not settled is \emph{what the grouping should optimise}. The natural
answer, and the one taken by the spatio-temporal clustering literature, is
proximity: group requests that are close in space and close in time. We show
that the second half of that instinct is backwards. A courier serves its group
one stop after another, so two requests that must both be served at ten o'clock
are not natural partners but competitors for the same courier-minute, while two
requests three hours apart are ideal partners. A clustering cost that rewards
temporal closeness therefore concentrates exactly the requests a courier cannot
serve. The effect is measurable and is not particular to our formulation: adding
a scaled time coordinate to $k$-means, the standard way to make a clustering
``spatio-temporal'', multiplies its tardiness by $3.5$ on instances where
spatial and temporal structure do not already agree, and no weight on that
coordinate recovers the loss at any setting we tried.

The fix is to stop describing a group geometrically. A group is good when its
requests \emph{chain}: when the courier can leave one and reach the next inside
its window. That is a property of a path, so we price a request against a
courier by the marginal cost of the routing subproblem: the increase in
travel plus lateness when that request is inserted into what the courier is
already doing. The cost is computed from the current assignment and then held
fixed while the assignment problem is solved, which keeps that problem a
well-structured integer program, and it is vehicle-aware in a way a centroid
cannot be: it knows where the courier is and when it becomes free.

What that correction buys is not punctuality but distance. Across a twentyfold
range in the penalty rate $\rho$, our solutions are consistently shorter than
the best clustering rule's and consistently a little later
(Section~\ref{sec:rho}). That follows from what chaining is: a group whose
requests chain is one whose courier reaches each next request without a long
empty run, so the travel between its stops is short. Punctuality is available to
either method once lateness is priced highly enough, and what separates them is
the distance each pays to reach it.

The problem this addresses is the pickup and delivery problem with time windows
(PDPTW) under conditions particular to courier dispatch. It generalises the
vehicle routing problem~\citep{dantzig1959truck, clarke1964scheduling} by pairing
each pickup with a specific delivery and restricting service at both to
prescribed windows~\citep{dumas1991pickup, savelsbergh1995general,
parragh2008surveyII}; applications span last-mile
delivery~\citep{agatz2011time, ulmer2018budgeting}, home health
care~\citep{mankowska2014home, NASIR2020113361} and ride-sharing
\citep{agatz2012optimization, qin2022reinforcement}. Terminology in this area is
not used consistently, so \tabref{tab:taxonomy} sets out the structural axes
that separate these problems and locates ours among them. Three entries in our
row are what make the clustering question different: windows are \emph{soft}, so
a bad grouping surfaces as cost rather than infeasibility; couriers start from
their own positions rather than a depot, so a group's cost depends on which
courier takes it; and routes are \emph{open}, so a group is a path rather than a
tour and its cost is not symmetric in the way a centroid assumes.

% ===== BEGIN TAXONOMY TABLE =====
% Structural taxonomy of pickup-and-delivery routing problems.
% Ported unchanged from the earlier version of the paper; the row for this
% paper still describes the setting correctly (soft windows, per-vehicle
% origins, open routes).
\begin{sidewaystable}[p]
\centering
\caption{Structural taxonomy of pickup-and-delivery routing problems. Commodity flow
follows the standard classification~\citep{berbeglia2007static, berbeglia2010dynamic}: $1$--$1$ denotes
commodities each having one origin and one destination, and $1$--$M$--$1$ commodities
delivered from a depot to customers and collected from customers to a depot. A double
dagger marks the row in which several pickups belong to one request: each commodity is
still $1$--$1$, since the goods collected at each pickup have one origin and one
destination. The third
class of that scheme, many-to-many, in which any location may be the origin or
destination of any commodity, does not arise among the variants compared here. A dagger
marks the degenerate case of $1$--$M$--$1$ in which no pickups occur, so that goods flow
only outward from the depot. ``Composition restriction'' records what limits the order in
which a route may interleave the requests it carries; \emph{none} means any order
consistent with precedence and capacity is permitted. Vehicle capacity is a constraint
throughout and is therefore not shown. PDP abbreviates pickup and delivery problem.}
\label{tab:taxonomy}
\renewcommand{\arraystretch}{1.15}
\resizebox{\textheight}{!}{%
\begin{tabular}{|l|c|c|c|l|c|c|c|l|}
\hline
\textbf{Problem variant} &
\textbf{Commodity} &
\textbf{Pickups} &
\textbf{Deliveries} &
\textbf{Composition} &
\textbf{Time} &
\textbf{Vehicle} &
\textbf{Route} &
\textbf{Representative} \\
 &
\textbf{flow} &
\textbf{per req.} &
\textbf{per req.} &
\textbf{restriction} &
\textbf{windows} &
\textbf{origins} &
\textbf{form} &
\textbf{work} \\
\hline\hline
Capacitated VRP (CVRP) &
$1$--$M$--$1^{\dagger}$ & 0 & 1 & none & none & single depot & closed &
\citep{dantzig1959truck, clarke1964scheduling} \\
\hline
VRP with time windows (VRPTW) &
$1$--$M$--$1^{\dagger}$ & 0 & 1 & none & hard & single depot & closed &
\citep{solomon1987algorithms} \\
\hline
VRP with backhauls (VRPB) &
$1$--$M$--$1$ & 0 or 1 & 1 or 0 & all deliveries before pickups & none or hard & single depot & closed &
\citep{osman2002reactive} \\
\hline
VRP with simultaneous P\&D (VRPSPD) &
$1$--$M$--$1$ & 1 & 1 & none & none or hard & single depot & closed &
\citep{parragh2008surveyI} \\
\hline\hline
Full-truckload PDP &
$1$--$1$ & 1 & 1 & one request at a time & hard & multiple depots & closed &
\citep{gronalt2003new} \\
\hline
PDP with LIFO loading (PDTSPL) &
$1$--$1$ & 1 & 1 & last-in-first-out unloading & none or hard & single depot & closed &
\citep{parragh2008surveyII} \\
\hline
PDP with time windows (PDPTW) &
$1$--$1$ & 1 & 1 & none & hard & single depot & closed &
\citep{dumas1991pickup, ropke2007models, ropke2009branch, baldacci2011exact} \\
\hline
Dial-a-ride problem (DARP) &
$1$--$1$ & 1 & 1 & maximum ride time per request & hard & single depot & closed &
\citep{CORDEAU2003579, cordeau2006branch} \\
\hline
Multi-pickup PDPTW (MPDPTW) &
$1$--$1^{\ddagger}$ & $\ge 1$ & 1 & none & hard & single depot & closed &
\citep{naccache2018multi, aziez2020exact} \\
\hline
Meal delivery routing (MDRP) &
$1$--$1$ & 1 & 1 & none & hard (deadline) & per vehicle & open &
\citep{reyes2018meal, yildiz2019provably, ulmer2021restaurant} \\
\hline\hline
\textbf{This paper} &
$1$--$1$ & 1 & 1 & none & \textbf{soft} & \textbf{per vehicle} & \textbf{open} &
--- \\
\hline
\end{tabular}%
}
\end{sidewaystable}
% ===== END TAXONOMY TABLE =====

\titledParagraph{Related work.}
Exact treatment of the PDPTW begins with column generation over a set-partitioning
formulation, embedded in branch-and-bound and priced by dynamic
programming~\citep{dumas1991pickup}; the window tightening introduced there is
still standard and we reuse it. Branch-and-cut formulations with families of
valid inequalities followed~\citep{ropke2007models, cordeau2006branch}, together
with branch-and-cut-and-price~\citep{ropke2009branch} and set partitioning with
dual ascent~\citep{baldacci2011exact}, which remain the strongest exact methods;
related exact work covers the capacitated VRP~\citep{baldacci2004exact} and
stochastic travel times~\citep{laporte1992vehicle}, a literature
surveyed in~\citep{GENDREAU19963}.
These solve instances of a few hundred requests, but take minutes to hours, and
an exact method interrupted early may return nothing feasible. A rolling horizon
allows seconds.

The field has therefore turned to heuristics: tabu search~\citep{nanry2000solving,
CORDEAU2003579}, building on tabu search for vehicle routing without
paired requests~\citep{gendreau1994tabu, cordeau2001tabu}, tabu-embedded simulated
annealing~\citep{li2001metaheuristic, li2003metaheuristic}, insertion and
route-building construction~\citep{potvin1993parallel, lu2006new,
mitrovic2004double}, and grouping genetic algorithms~\citep{pankratz2005grouping};
multi-objective variants are surveyed in~\citep{jozefowiez2008multi}. The
strongest current methods are large-neighbourhood searches: adaptive large
neighbourhood search~\citep{ropke2006adaptive} improved a majority of best-known
solutions on the Li \& Lim instances~\citep{SINTEFLiLimBenchmark}, and
\citet{curtois2018large} combined local search, large neighbourhood search and
guided ejection search to improve them further. A parallel line learns
construction heuristics directly~\citep{NIPS2015_29921001, bello2016neural,
nazari2018reinforcement, kool2019attention}, including for pickup
and delivery~\citep{li2021heterogeneous, li2021learning}; these are fast at
inference but require training per instance distribution and give no feasibility
guarantee under capacity and window constraints. Deep policy dynamic
programming~\citep{kool2022deep} uses a learned model differently, to restrict a
dynamic program that extends only feasible partial routes, so it does respect
time windows, at a higher inference cost than direct construction.

Two literatures bear directly on our setting. Requests spanning several merchants
(the multi-pickup problem~\citep{naccache2018multi, aziez2020exact}) form a
different combinatorial object from the paired requests studied here. Meal
delivery matches our operational conditions closely: the routing problem was
introduced in~\citep{reyes2018meal} with a rolling-horizon matching algorithm, the
static variant solved by simultaneous column and row generation on industry data
in~\citep{yildiz2019provably}, random meal ready times added
in~\citep{ulmer2021restaurant} and multi-restaurant orders
in~\citep{steever2019dynamic}. In all of these, as here, couriers are dispatched
from wherever their previous assignment left them.

The present work belongs to the cluster-first, route-second tradition introduced
in~\citep{fisher1981generalized}, where customers are assigned to vehicles by a
generalised assignment problem whose objective approximates routing cost, after
which each vehicle is routed independently. The assignment step is tractable
due to~\citep{bradley2000constrained}, which showed that $k$-means under
cluster-size constraints is a minimum-cost flow problem solvable as a linear
program. Within routing, $k$-means and its relatives have been used to assign
customers to depots~\citep{dupas2024clustering}, balance workload across delivery
areas~\citep{mo2024multi}, and support cold-chain
distribution~\citep{shi2024optimization}, facility location~\citep{zhou2020joint},
the VRP with time windows~\citep{le2022clustering} and two-echelon
routing~\citep{wang2024hybrid}; hybrid $k$-means and genetic schemes appear
in~\citep{fatemi2022hybrid, wang2021multi}, and a three-stage heuristic for the
capacitated VRP with time windows in~\citep{akeb2015three}. Algorithmic
foundations are treated in~\citep{hruschka2009survey, selim1984k}.

Three features of this stream motivate the present work. \emph{First, the number
of groups is tied to the fleet.} A cluster \emph{is} a vehicle's workload, so the
two are in one-to-one correspondence~\citep{fisher1981generalized,
akeb2015three, dupas2024clustering, le2022clustering}; every vehicle is
implicitly dispatched and fleet size cannot be traded against travel, even when a
courier carries a fixed cost. \emph{Second, the grouping objective is a geometric
proxy for a routing cost, and is not validated against it.} A request is scored by
its distance to a centroid, a quantity that summarises a whole workload by one
point and is never compared with what the routing subproblem would actually
charge. \emph{Third, groups are formed on proximity, and the standard remedy for
time windows is to extend that proximity to the time axis}, by a spatio-temporal
embedding or a distance from a request's window to a cluster's mean time. We
show in Section~\ref{sec:cost} that this is the wrong relation: under sequential
service, requests due at the same moment compete rather than belong together, so
a temporal proximity term concentrates precisely the requests one courier cannot
serve. The three are addressed by one change: pricing a request against a
courier by the marginal cost of the routing subproblem, in a program that also
decides which couriers to open.

\smallskip
\titledParagraph{Contributions.}
\begin{enumerate}[label=(\roman*),leftmargin=*,noitemsep]
\item We show by measurement that temporal proximity is the wrong quantity to
cluster on under sequential service, and we replace it with the marginal cost of
the induced routing subproblem (Section~\ref{sec:cost}).
\item We make the grouping dispatch-aware, formulating it as one integer program
that assigns requests to couriers and simultaneously decides which couriers to
open, and we keep that program linear in the number of requests with candidate
lists (Section~\ref{sec:assign}).
\item We build two cheap surrogates for the subproblem cost, measure their
\emph{rank} fidelity against the exact router rather than assume it, and use
both in a two-stage descent, the faster screening every candidate move and the
sharper re-scoring only the promising ones
(Sections~\ref{sec:surrogate}--\ref{sec:descent}).
\item We derive a parallel form of the assignment step in which each courier
solves its own subproblem, coupled only by a multiplier on each request, and we
report what that form costs as well as what it saves
(Section~\ref{sec:lagrangian}).
\item We evaluate against six published clustering rules, in their published
form and strengthened with our own improvement step, over three problem sizes,
two alignment regimes and benchmark batches we did not design, scaling to 1000
requests and 300 couriers, and we mark the boundary of the claim by removing the
dispatch charge (Section~\ref{sec:sims}).
\end{enumerate}

%%%%%%%%%%%%%%%%%%%%%%%%%%%%%%%%%%%%%%%%
%%%%        PROBLEM STATEMENT       %%%%
%%%%%%%%%%%%%%%%%%%%%%%%%%%%%%%%%%%%%%%%

\section{Problem Statement}\label{sec:problemFormulation}

\subsection{Data}\label{sec:data}
An instance consists of requests $\mathcal{R}$ and couriers $\mathcal{K}$.
Request $r\in\mathcal{R}$ is a pair $(p_r,d_r)$ of a pickup and a delivery;
writing $\mathcal{P}=\{p_r\}$ and $\mathcal{D}=\{d_r\}$, the stops are
$\mathcal{N}=\mathcal{P}\cup\mathcal{D}$. Stop $i$ has a service duration
$S_i\ge 0$ and a window $[e_i,l_i]$ within which service should \emph{begin}.
Courier $k$ has capacity $Q_k$, an origin $o_k$, the location it starts from,
and a ready time $\theta_k$, the time it becomes available. Travel times $T_{ij}$ satisfy the triangle inequality.

Each request occupies one unit of capacity, so $Q_k$ counts requests rather than
weight: what binds in meal and parcel delivery is how many orders fit in the
bag. \emph{Capacity is a bound on the total number of requests a courier
receives}, $|\mathcal{R}_k|\le Q_k$, not on the number it holds at one instant.
The two readings differ, and the choice is forced by the decomposition rather
than by the application. A cluster-first method commits a whole group to a
courier \emph{before} any sequence exists, so at the moment the capacity
constraint is imposed there is nothing to count concurrency on; only the total
is known. The total-count reading is the stronger of the two, since a route never
holds more than it is given, so a solution feasible under it is feasible under
the simultaneous-load reading as well, and every method in
Section~\ref{sec:sims}, including ALNS, is held to it so that no method is
allowed to close a courier the others may not. One consequence is used
repeatedly in Section~\ref{sec:sims}: capacity alone forces at least
$\lceil|\mathcal{R}|/\max_k Q_k\rceil$ couriers, a floor no method can undercut.

Two modelling choices follow dispatch practice rather than the classical PDPTW.
\emph{Routes are open paths}: couriers are dispatched on a rolling horizon and
end wherever the last delivery leaves them, which becomes the origin of the next
round. \emph{Time windows are soft}: a batch released at once often admits no
schedule meeting every window, and a dispatcher cannot abandon the batch, so
violations are priced at a rate $\rho$ rather than forbidden. A consequence we
rely on is that no grouping can make a subproblem infeasible: a bad grouping
surfaces as cost, which is a signal an improvement step can act on.

\subsection{Solutions and objective}
A solution assigns each request to exactly one courier and gives each courier an
open path $\pi_k$ from its origin through the stops of its requests, subject to pairing
(a request's two stops are served by the same courier), precedence (pickup
before delivery), capacity, and time consistency. Writing $b_i$ for the time
service begins at stop $i$,
\begin{equation}\label{eq:penalties}
t^{-}_i=\bigl[e_i-b_i\bigr]^{+},\qquad t^{+}_i=\bigl[b_i-l_i\bigr]^{+},
\end{equation}
are the \emph{earliness} and the \emph{tardiness} at stop $i$, the amounts by
which service begins before the window opens and after it closes. With
$u_k\in\{0,1\}$ indicating that courier $k$ is dispatched and
$\operatorname{tt}(\pi_k)$ the travel time of its path,
\begin{equation}\label{eq:globalobj}
\min\;\;
\underbrace{\sum_{k\in\mathcal{K}}\operatorname{tt}(\pi_k)}_{\text{travel}}
\;+\;\rho\underbrace{\sum_{i\in\mathcal{N}}\bigl(t^{+}_i+t^{-}_i\bigr)}_{\text{service quality}}
\;+\;\varphi\underbrace{\sum_{k\in\mathcal{K}}u_k}_{\text{fleet}} .
\end{equation}
The fleet term is what makes the dispatched fleet a decision, and we call
$\varphi$ the \emph{dispatch charge}, the cost of putting one more courier on the
road. Without it an
extra courier can only reduce travel and lateness, so every courier is used at
optimality and the question does not arise; with it, the model must weigh a
courier's fixed cost against the travel and punctuality it buys. We report both
regimes.

\begin{remark}[Earliness]\label{rem:earliness}
With earliness and tardiness priced at the same rate, the earliest feasible
schedule $b_i=\max\{\text{arrival}_i,e_i\}$ is optimal for any fixed sequence.
Two deviations are available and neither pays. Delaying service beyond
$\max\{\text{arrival}_i,e_i\}$ leaves $t^-_i=0$ untouched and can only add
tardiness, at stop $i$ and at every stop after it. Serving \emph{before} $e_i$
is permitted by~\eqref{eq:penalties}, and buys a reduction in downstream
tardiness, but it does so by converting that tardiness into earliness one unit
for one; at a common rate the exchange is exactly neutral. The earliest feasible
schedule therefore attains the optimum, though not uniquely, and it is the
optimal schedule at which $t^-_i=0$ at every stop. Sequencing therefore reduces to a shortest path
over subsets, which is what makes exact routing affordable at the sizes the
decomposition produces. Asymmetric pricing $\rho^-<\rho^+$ breaks this and
requires the schedule to be solved as a small linear program.
\end{remark}

A consequence worth stating, since it makes one term of~\eqref{eq:globalobj}
inert: under this scheduling rule $t^-_i=0$ at every stop, so the earliness term
never contributes and every table in Section~\ref{sec:sims} reports tardiness alone. We keep the
term in the objective rather than deleting it because it is what distinguishes
this model from one that prices only lateness, and because the moment
$\rho^-<\rho^+$, the realistic pricing for courier work and the extension
Remark~\ref{rem:earliness} points at, the term becomes active and the
scheduling rule, not the objective, is what must change.

\subsection{Problem difficulty and the case for decomposition}
The PDPTW contains the VRP and is NP-hard~\citep{lenstra1981complexity}, but that
is not the informative statement. A monolithic formulation carries
$\Theta(|\mathcal{R}|^2|\mathcal{K}|)$ binaries because it does not know the
assignment. Fixing the assignment eliminates that difficulty exactly: the
courier index becomes a label on a set of requests, pairing and
single-service constraints
are enforced by the partition, and the problem separates into $|\mathcal{K}|$
independent sequencing problems on $2|\mathcal{R}_k|$ stops. Sequencing is not
eliminated but shrunk, and because its effort grows steeply in the number of
stops, keeping $|\mathcal{R}_k|$ small is what makes exact routing possible. The
decomposition's task is therefore to produce groups that are small, feasible,
and \emph{chainable}, the last of these being the subject of this paper.

%%%%%%%%%%%%%%%%%%%%%%%%%%%%%%%%%%%%%%%%
%%%%             METHOD             %%%%
%%%%%%%%%%%%%%%%%%%%%%%%%%%%%%%%%%%%%%%%

\section{Method}\label{sec:method}

\subsection{Overview}
\begin{figure}[!htbp]\centering
\centering
\includegraphics[width=0.40\textwidth]{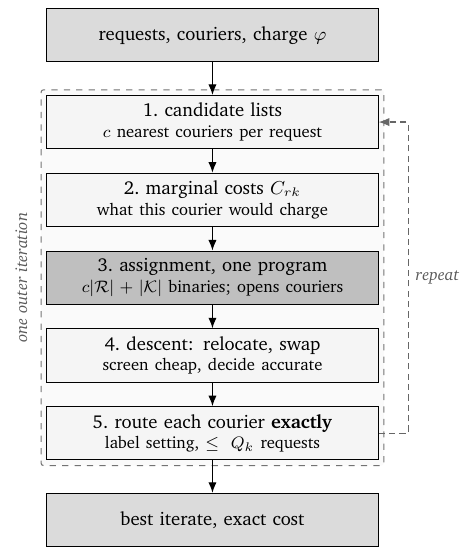}
\caption{The proposed method end to end. The dashed region is one outer iteration:
steps~1--5 run in order and then repeat, and what our proposed method returns is the
cheapest iterate any pass produced, not the last
(Section~\ref{sec:guarantees}). Shading marks how a step behaves when cores are
added: light steps are the ones whose expensive work decomposes by
courier, the dark step is the one integer program that does not decompose at
all, because each request must be taken exactly once; the light steps still
carry synchronisation points, which Section~\ref{sec:parallel} sets out. Section~\ref{sec:scaling} measures what each step costs. Time enters
only through step~2: what lateness a courier would actually incur, never how
close two requests are in time.}
\label{fig:pipeline}
\end{figure}

\begin{figure}[!htbp]\centering
\includegraphics[width=\textwidth]{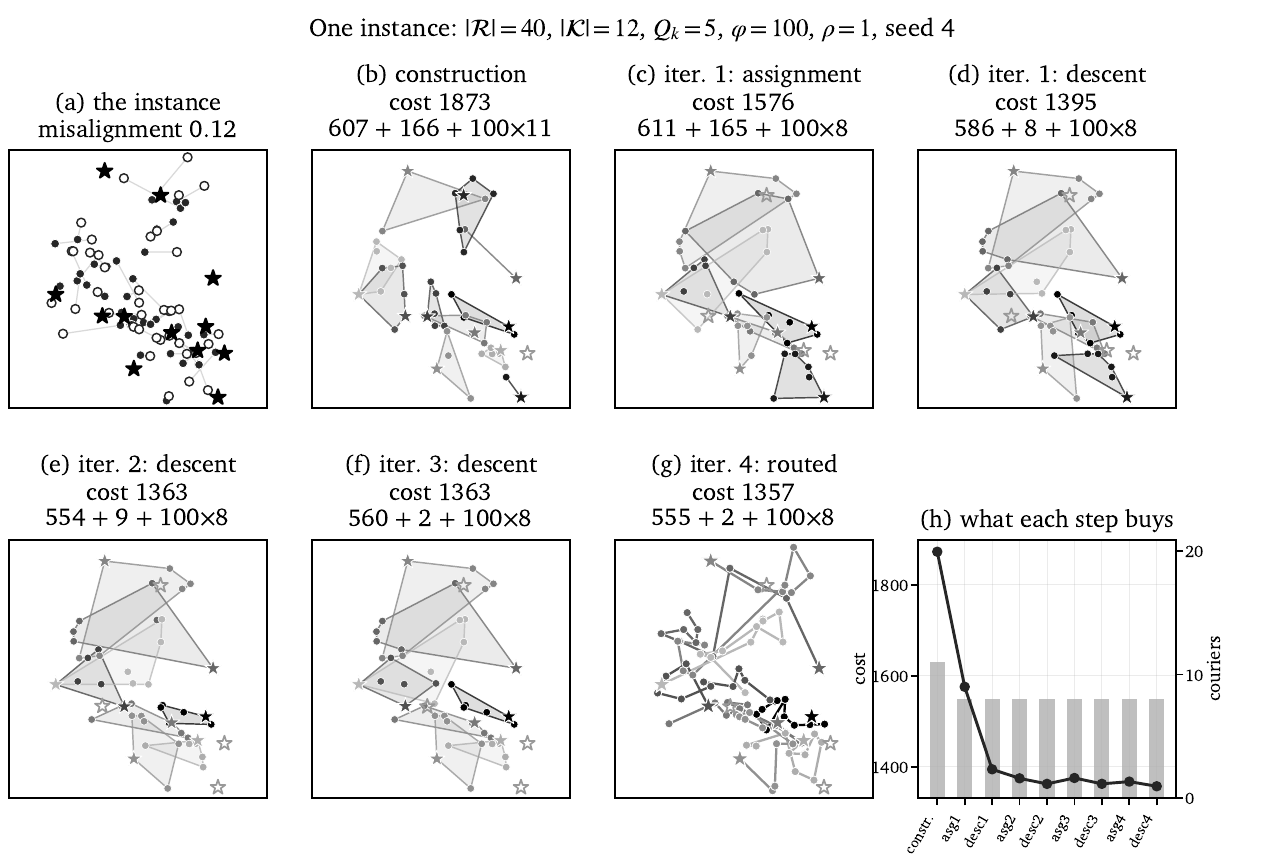}
\caption{The proposed method running on one instance, drawn from a trace the solver
emits rather than reconstructed, so each panel is a state the algorithm actually
passed through. Panel (a) is the instance as posed; in it and in every map
panel a filled circle is a pickup and an open one a delivery. In (b)--(g) a
shaded region
encloses each courier's group together with where that courier starts, and
hollow stars mark couriers left idle. Panels titled \emph{descent} are the
state after that iteration's improvement step; (g) additionally shows the exact routes,
whose nodes are the pickups and deliveries the courier visits rather than the
group midpoints of the earlier panels. Below each cost are its three terms in
the order of~\eqref{eq:globalobj}: travel, lateness, and $\varphi$ times the
couriers dispatched.
The initial construction opens eleven couriers and the first assignment program closes
three of them at once, which is the step no published clustering rule performs.
Later iterations move a few requests at a time: the fleet is settled early and
the remaining gain is in which courier holds what. Panel (h) reads the same
history as numbers, with cost on the left axis and the dispatched fleet on the
right.}
\label{fig:steps}
\end{figure}

Our proposed method repeats five steps, shown in \figref{fig:pipeline}, and every
reported cost comes from the exact router. One repetition of the five is a
\emph{pass}, the outer loop's iteration, and the word is used in that sense
throughout. \figref{fig:steps} follows one instance through
four passes, from the construction to the routed solution.

\begin{enumerate}[label=\arabic*.,leftmargin=*,noitemsep]
\item \textbf{Candidate lists.} For each request keep the $c$ couriers whose
current work is nearest, rebuilt from the current solution.
\item \textbf{Marginal costs.} For each allowed (request, courier) pair compute
$C_{rk}$, the increase in that courier's routing cost if it took the request.
\item \textbf{Assignment.} Solve one integer program that assigns every request
to a courier and decides which couriers to open.
\item \textbf{Descent.} Improve the assignment by relocate and swap moves,
screened with a cheap surrogate and decided with an accurate one.
\item \textbf{Routing.} Route every dispatched courier exactly, and keep the
best iterate seen.
\end{enumerate}

\subsection{The assignment cost}\label{sec:cost}
Let $\mathcal{R}_k$ be the requests currently on courier $k$ and $J(\mathcal{S},k)$
the cost of serving a set of requests $\mathcal{S}$ on courier $k$: travel plus $\rho$ times
lateness, scheduled by the earliest-feasible rule of
Remark~\ref{rem:earliness}. The assignment cost is the marginal cost
\begin{equation}\label{eq:marginal}
C_{rk}=
\begin{cases}
J(\mathcal{R}_k\cup\{r\},k)-J(\mathcal{R}_k,k), & r\notin\mathcal{R}_k,\\[2pt]
J(\mathcal{R}_k,k)-J(\mathcal{R}_k\setminus\{r\},k), & r\in\mathcal{R}_k,
\end{cases}
\end{equation}
so that members and non-members are priced on the same scale. The two branches
are one quantity: writing $\Delta(r\mid\mathcal{S})=J(\mathcal{S}\cup\{r\},k)
-J(\mathcal{S},k)$ for what $r$ adds to a set of requests $\mathcal{S}$ it does
not belong to, both read $C_{rk}=\Delta(r\mid\mathcal{R}_k\setminus\{r\})$, since
$\mathcal{R}_k\setminus\{r\}=\mathcal{R}_k$ when $r$ is not already there. The
marginal cost $C_{rk}$ is computed from the current assignment and then held fixed while the assignment problem is solved,
which keeps that problem a fixed-cost integer program.

\titledParagraph{Why not a spatio-temporal distance.} The alternative, and the
standard choice, is to embed requests in a joint space and group by proximity:
a spatial term plus a temporal term such as the squared distance from a
request's time interval to a group's mean time. Under sequential service this is
the wrong object. A courier visits its requests one after another, so two
requests whose windows coincide compete for the same courier-minute, and two
requests far apart in time are precisely the pair one courier can serve
comfortably. Minimising a temporal distance concentrates the competing requests.

The same error can appear in a subtler form. A natural remedy is to charge each
pair of requests the courier cannot chain. Writing
$\operatorname{slack}(r\!\to\!r')=(e_{p_{r'}}-l_{d_r})-T(d_r,p_{r'})$ for the
spare time between delivering $r$ and collecting $r'$, the quantity
$[-\operatorname{slack}(r\!\to\!r')]^{+}$ is how far that order falls short of
feasible. A grouping cost has to charge the pair rather than the ordered pair, so
the two directions are combined, either as
\begin{equation}\label{eq:chaincharge}
\begin{aligned}
g_{\mathrm{sum}}(r,r')&=[-\operatorname{slack}(r\!\to\!r')]^{+}
+[-\operatorname{slack}(r'\!\to\!r)]^{+},\\
g_{\min}(r,r')&=\min\bigl\{[-\operatorname{slack}(r\!\to\!r')]^{+},\;
[-\operatorname{slack}(r'\!\to\!r)]^{+}\bigr\},
\end{aligned}
\end{equation}
and requests are then grouped to minimise the total charge inside each group. But
a courier \emph{chooses} the order, so a pair is unchainable only if \emph{both}
orders fall short, which is what $g_{\min}$ charges and $g_{\mathrm{sum}}$ does
not: $g_{\mathrm{sum}}$ charges a pair whenever \emph{either} order fails, and so
charges most for the pairs that chain most easily. On
our generated instances $g_{\mathrm{sum}}$ correlates $+0.95$ with the
temporal separation $|\tau_r-\tau_{r'}|$, writing
$\tau_r=\tfrac{1}{2}(e_{p_r}+l_{d_r})$ for the midpoint of the span over which
request $r$ must be served. It is, to a good approximation, a measure
of how far apart in time two requests are. Section~\ref{sec:sims} reports what
each of these choices costs.

\subsection{The assignment program}\label{sec:assign}
With costs fixed, let $x_{rk}\in\{0,1\}$ assign request $r$ to courier $k$,
$u_k\in\{0,1\}$ open courier $k$, and $\mathcal{C}(r)\subseteq\mathcal{K}$ be
the couriers $r$ may be assigned to, the \emph{candidate list} defined
after~\eqref{eq:cap}:
\begin{subequations}\label{eq:assignprog}
\begin{align}
\min\quad & \sum_{r\in\mathcal{R}}\sum_{k\in\mathcal{C}(r)} C_{rk}\,x_{rk}
          \;+\;\varphi\sum_{k\in\mathcal{K}} u_k \label{eq:assign}\\
\text{s.t.}\quad & \sum_{k\in\mathcal{C}(r)} x_{rk}=1, && \forall r\in\mathcal{R},
          \label{eq:cover}\\
& \sum_{r:\,k\in\mathcal{C}(r)} x_{rk}\le Q_k\,u_k, && \forall k\in\mathcal{K},
          \label{eq:cap}\\
& x_{rk},u_k\in\{0,1\}. \nonumber
\end{align}
\end{subequations}
Constraint~\eqref{eq:cover} is the only one that couples couriers, and it is
what makes~\eqref{eq:assignprog} a single program rather than $|\mathcal{K}|$
independent ones. The fleet decision sits in the $u_k$, priced by $\varphi$,
which is what allows our proposed method to trade couriers against travel; no published
clustering rule we are aware of can make that trade, because each partitions into
a fixed number of groups.

\begin{remark}[Fixed charge and integrality]\label{rem:fixedcharge}
With $\varphi=0$ and $u_k$ dropped, \eqref{eq:assignprog} is a
transportation problem: the constraint matrix is totally unimodular, the
relaxation is integral, and the assignment is obtained by linear programming.
The fixed charge $\varphi u_k$ destroys that property, which is why the fleet
decision is the expensive part of the program and, as
Section~\ref{sec:lagrangian} shows, the part that resists decomposition.
\end{remark}

\titledParagraph{Candidate lists.} A request is rarely served by a courier on
the far side of the city, so we keep for each request $r$ a set
$\mathcal{C}(r)$ of the $c$ nearest couriers. What that distance is measured to
matters, because a courier has no fixed location: we take its position to be the
mean of the midpoints of the requests currently assigned to it, so a courier
counts as near a request when the work it already carries is near that request,
not when its starting point happens to be. The lists are therefore rebuilt each
outer iteration from the current solution, and the incumbent courier is always
kept in $\mathcal{C}(r)$ so that solution stays feasible. A courier holding
nothing has no current work to measure a distance from, so its own origin $o_k$
stands in: without that an empty courier would enter no candidate list, the
program could never open it, and the fleet could only shrink. The restriction is
a heuristic. When $|\mathcal{K}|\le c$ it is vacuous and the program is the
dense one, and a distant courier can be the right choice whenever the windows
are loose enough to absorb the drive. But it is what makes the program small.
It reduces the binaries from $|\mathcal{R}||\mathcal{K}|$ to
$c|\mathcal{R}|+|\mathcal{K}|$ and the marginal-cost evaluations from
$|\mathcal{R}||\mathcal{K}|$ to $c|\mathcal{R}|$, both linear in
$|\mathcal{R}|$ for fixed $c$. 

\subsection{Surrogates with measured fidelity}\label{sec:surrogate}
The exact router of Section~\ref{sec:routing} is far too expensive to call
inside the search, because it is not consulted once but
continually. The marginal cost of~\eqref{eq:marginal} is wanted for every
allowed (request, courier) pair each time the candidate lists are rebuilt, and
again for every move the descent of Section~\ref{sec:descent} weighs: a single
solve at $|\mathcal{R}| = 200$ asks for $J(\cdot,k)$ about $400{,}000$ times.
Answering all of them exactly would turn the $18$\,s our proposed method takes at that
size into hours (Section~\ref{sec:scaling}). Two cheap surrogates are available, and
what matters is not how close their \emph{value} is but whether they \emph{rank}
two candidate request sets the way the exact router would. \tabref{tab:surrogate}
measures both against the exact router over 120 random request sets.

\begin{table}[!htbp]\centering
\begin{minipage}{0.80\textwidth}
\centering
\caption{Three ways to cost one courier's request set, over 120 random sets.
What a descent needs is the \emph{rank}, not the value.}
\label{tab:surrogate}
\small\begin{tabular*}{\textwidth}{@{\extracolsep{\fill}}l r r r@{}}
\toprule
 & ratio & rank corr. & time\\
\midrule
stops in time-window order       & $2.418$ & $0.949$ & ${\sim}1\,\mu$s\\
cheapest insertion in time order & $1.043$ & $0.990$ & $709\,\mu$s\\
exact label-setting router       & $1.000$ & $1.000$ & $31.6\,$ms\\
\bottomrule
\end{tabular*}
\end{minipage}
\end{table}

Ordering stops by time window ignores geography and zig-zags, overstating cost
by a factor of $2.4$; cheapest insertion lands within $4.3\%$ of exact at one
forty-fourth of the cost. We use the first to screen and the second to decide,
and call them the \emph{window-order} and \emph{insertion} surrogates
from here on.

\subsection{Assignment descent}\label{sec:descent}
Write $\hat{J}$ for the insertion surrogate of Section~\ref{sec:surrogate}, with
$\hat{J}(\emptyset,k)=0$, and let
\begin{equation}\label{eq:descentobj}
\Phi(a)\;=\;\sum_{k\,:\,\mathcal{R}_k\neq\emptyset}
          \bigl[\hat{J}(\mathcal{R}_k,k)+\varphi\bigr]
\end{equation}
be the objective of~\eqref{eq:globalobj} under an assignment $a$, with the exact
$J$ replaced by $\hat{J}$. The dispatch charge is carried inside $\Phi$: without
it, emptying a courier would look free and the descent could not see the one
saving the fleet decision exists to find.

Two move classes are considered. \emph{Relocate} moves a request $r$ to a
courier in $\mathcal{C}(r)$ with room; \emph{swap} exchanges $r_1$ and $r_2$
when each is a candidate for the other's courier, which is what keeps the number
of swap moves linear in $|\mathcal{R}|$ rather than quadratic. Both classes are
enumerated in full, not sampled: no step of the descent is random. Each move is first scored
with the window-order surrogate, the $\tau$ most promising by that score are
re-scored with $\hat{J}$, and the one whose $\Delta\Phi$ is most negative is
applied, provided $\Delta\Phi<-\epsilon$. A sweep of the move set that produces
no such move ends the descent. We use $\tau=8$ and $\epsilon=10^{-9}$.

Termination and the absence of cycling both follow from this. Every applied move
lowers $\Phi$ by more than $\epsilon$; $\Phi$ depends on the assignment alone,
not on the path taken to it; and the assignments form a finite set. So no
assignment can recur, so in particular the descent cannot oscillate between two
solutions, and it halts after finitely many moves. The margin $\epsilon$ is
what rules out an endless sequence of improvements too small to accumulate in
floating point. Note what is \emph{not} claimed: the fixed point is one of
$\Phi$, not of the exact objective, and it is a local minimum with respect to
these two move classes only.

\subsection{Routing the induced subproblem, exactly}\label{sec:routing}
Once the assignment is fixed, each dispatched courier faces a routing problem of
its own, and those problems share nothing: no variable, no constraint, and no
decision. This subsection states that subproblem, gives the algorithm that
solves it, and shows the algorithm is exact.

\titledParagraph{The subproblem.} Fix a courier $k$ and let
$\mathcal{R}_k\subseteq\mathcal{R}$ be the requests assigned to it. Define
\begin{equation}\label{eq:subsets}
\begin{aligned}
\mathcal{P}_k &= \{p_r : r\in\mathcal{R}_k\}, &
\mathcal{D}_k &= \{d_r : r\in\mathcal{R}_k\},\\
\mathcal{N}_k &= \mathcal{P}_k\cup\mathcal{D}_k, &
\mathcal{V}_k &= \mathcal{N}_k\cup\{o_k,\varepsilon_k\},
\end{aligned}
\end{equation}
where $o_k$ is the courier's origin and $\varepsilon_k$ an artificial terminal
with $T_{i\varepsilon_k}=0$ for all $i$ and no outgoing arcs. A route is a
Hamiltonian path on $\mathcal{V}_k$ from $o_k$ to $\varepsilon_k$: an order
$\pi_k=(i_0,i_1,\dots,i_m)$ with $i_0=o_k$, $m=|\mathcal{N}_k|=2|\mathcal{R}_k|$
and every stop of $\mathcal{N}_k$ appearing once. Because the closing arc to
$\varepsilon_k$ is costless the courier is never charged for returning anywhere,
which is what makes the route open.

An order fixes the schedule. Writing $b_i$ for the time service begins at stop
$i$, as in Section~\ref{sec:problemFormulation}, the earliest-feasible rule of
Remark~\ref{rem:earliness} gives
\begin{equation}\label{eq:sched}
b_{i_0}=\theta_k,\qquad
b_{i_q}=\max\bigl(b_{i_{q-1}}+S_{i_{q-1}}+T_{i_{q-1}i_q},\;e_{i_q}\bigr),
\quad q=1,\dots,m,
\end{equation}
with $S_{o_k}=0$. The subproblem is then to choose the order:
\begin{equation}\label{eq:subobj}
\min_{\pi_k}\;\;\sum_{q=1}^{m}T_{i_{q-1}i_q}
\;+\;\rho\sum_{q=1}^{m}\bigl[b_{i_q}-l_{i_q}\bigr]^{+}
\qquad\text{s.t. } p_r \text{ precedes } d_r \text{ in }\pi_k,\;
\forall r\in\mathcal{R}_k,
\end{equation}
and a non-empty $\mathcal{R}_k$ contributes a further $\varphi$. Only tardiness
appears because~\eqref{eq:sched} leaves $t^-_i=0$ at every stop, by
Remark~\ref{rem:earliness}; pairing, single service and capacity are not
constraints here but properties of $\mathcal{R}_k$, which the assignment has
already settled.

Three difficulties the assignment removes are worth naming, because they are what
make this subproblem small. There is \emph{no courier index}: variables are indexed by
arc and node only, so the $|\mathcal{K}|$-fold replication of timing and load
variables disappears. The \emph{coupling constraints vanish}: pairing, single
service and same-courier assignment are enforced by the partition rather than by
constraints, and only precedence survives, as a two-variable relation. And the
subproblems are \emph{independent}, so they may be solved in any order or
concurrently.

\titledParagraph{Earliest service times.} Soft windows leave the
subproblem with no arcs to eliminate, because nothing in~\eqref{eq:sched}
confines $b_i$ to $[e_i,l_i]$: a late service is charged, not forbidden, so the
window endpoints are not themselves bounds on $b_i$ and no arc is rendered
infeasible by lateness.
What does bound $b_i$ is the earliest the courier could arrive, giving
\begin{equation}\label{eq:earliest}
\underline{b}_i=
\begin{cases}
\theta_k+T_{o_k i}, & i\in\mathcal{P}_k,\\[2pt]
\theta_k+T_{o_k p_r}+S_{p_r}+T_{p_r d_r}, & i=d_r\in\mathcal{D}_k,
\end{cases}
\end{equation}
since a pickup cannot be served before the courier can reach it, and a delivery
not before its own pickup has been served and traversed. They enter the
search directly, as the arrival times from which~\eqref{eq:labelext} extends.

\titledParagraph{Label setting.} We do not solve the subproblem as a
mixed-integer program, and Remark~\ref{rem:earliness} is the reason. For a fixed
order of the stops the \emph{optimal} service times are not a further decision
but a formula, $b_i=\max\{\text{arrival}_i,e_i\}$: waiting never pays when
earliness and tardiness cost the same. Choosing \emph{when} to serve therefore
drops out of the subproblem and only the order remains, leaving a pure
sequencing problem with no continuous variables for a mixed-integer program to
carry.

The same fact is what allows a label, defined next, to be a pair of numbers
rather than a function. Under asymmetric pricing $\rho^-<\rho^+$ the best completion of a
partial route would depend on the whole trade-off between arriving early and
arriving late, so a label would have to carry cost as a function of arrival
time; the dominance argument of Proposition~\ref{prop:routerexact} compares two
scalars and would not survive that change.

A \emph{state} is a pair $(U,i)$: $U\subseteq\mathcal{N}_k$ is the set of stops
already served and $i\in U$ is the one served last. A \emph{label} on that state
is a pair $(c,t)$, where $t$ is the time service begins at $i$ and $c$ is the
cost accumulated so far, travel plus $\rho$ times the lateness incurred at the
stops of $U$. The search starts from the single label $(0,\theta_k)$ on
$(\{o_k\},o_k)$ and extends one stop at a time. Extending a label $(c,t)$ at
$(U,i)$ to a stop $j\notin U$ gives
\begin{equation}\label{eq:labelext}
t' = \max\bigl(t+S_i+T_{ij},\,e_j\bigr),\qquad
c' = c+T_{ij}+\rho\bigl[t'-l_j\bigr]^{+},
\end{equation}
which is~\eqref{eq:sched} applied one stop further on, accumulating the two sums
of~\eqref{eq:subobj} as it goes: $t'$ is the service time the order would give
$j$, and the bracket is the tardiness that order would pay there. A label is
therefore a partial route priced exactly as~\eqref{eq:subobj} prices a whole
one. The extension is allowed only when $j$ is \emph{reachable}: a delivery
$d_r$ requires $p_r\in U$. Capacity needs no check, since
$|\mathcal{R}_k|\le Q_k$ holds for the whole set of requests by construction. A route is
complete when $U=\mathcal{N}_k$, and the answer is the least $c$ among the
labels there, the closing arc to $\varepsilon_k$ being free.

What keeps this affordable is \emph{dominance}. Two labels on the same state are
comparable, and $(c,t)$ dominates $(c',t')$ when $c\le c'$ and $t\le t'$; the
dominated one is discarded. Both components are needed. Keeping only the
cheapest label would be wrong, because a cheaper label may arrive later and pay
for that lateness at every stop still to come; keeping only the earliest would
be wrong for the mirror reason. What survives on each state is its Pareto set.
Dominance is also the only step that discards anything, so whether the router is
exact turns on it alone.

\begin{proposition}[The router is exact]\label{prop:routerexact}
Label setting with the extension~\eqref{eq:labelext} and the dominance rule above
returns a minimiser of~\eqref{eq:subobj}.
\end{proposition}

\begin{proof}
Fix a state $(U,i)$ and let $g(U,i;c,t)$ be the least total cost of completing a
route from it, given arrival with cost $c$ at time $t$. By~\eqref{eq:labelext}
every later cost is non-decreasing in the incoming $c$ and every later service
time is non-decreasing in the incoming $t$, and lateness $[\,\cdot\,]^{+}$ is
non-decreasing in its argument; so $g$ is non-decreasing in both $c$ and $t$.
Hence if $(c,t)$ dominates $(c',t')$, every completion available to the second is
available to the first at no greater cost, and discarding the dominated label
removes no optimal route. Since every route from $o_k$ through
$\mathcal{N}_k$ corresponds to exactly one sequence of extensions, and no
extension that could lie on an optimal route is pruned (precedence
excludes only orderings that violate $p_r$ before $d_r$, which no feasible route
uses, and dominance is the only other step that discards anything), the
surviving label of least cost at $U=\mathcal{N}_k$ is optimal.
Termination is immediate: $U$ grows by one stop at every extension.
\end{proof}

\noindent Nothing in this argument depends on how many requests the courier
holds: label setting is exact at any $|\mathcal{R}_k|$. What grows with
$|\mathcal{R}_k|$ is the running time. A state is a visited set and a last stop,
so there are $O(2^{2|\mathcal{R}_k|}|\mathcal{R}_k|)$ of them, and on random
request sets the router takes $5.6$\,ms at $|\mathcal{R}_k|=5$, $0.74$\,s at $8$ and
$18$\,s at $10$, four to five times longer for each request added.

Our implementation therefore runs the exact router up to eight requests and
falls back to cheapest insertion improved by or-opt and 2-opt beyond it. That
fallback never fires in this paper, because capacity bounds the requests a courier holds
before the limit does: $|\mathcal{R}_k|\le Q_k=5$ throughout
Section~\ref{sec:sims}, so the largest set ever routed is five requests, and no
reported number rests on a heuristic route. This is what the decomposition buys.
It does not make routing easy; it keeps $|\mathcal{R}_k|$ in the range where an
exact router is affordable at all.

\begin{proposition}[Validity of the decomposition]\label{prop:decompvalid}
Let $\{\mathcal{R}_k\}_{k\in\mathcal{K}}$ partition $\mathcal{R}$ with
$|\mathcal{R}_k|\le Q_k$ for every $k$, and let $\pi_k$ be a feasible route for
each $k$. Then $\{\pi_k\}$ is feasible for the problem of
Section~\ref{sec:problemFormulation}, and its objective~\eqref{eq:globalobj}
equals the sum of the subproblem costs plus $\varphi$ times the number of
non-empty parts. Consequently $\sum_k z^{\star}_k \ge z^{\star}$, where
$z^{\star}_k$ is the optimal value of subproblem $k$ and $z^{\star}$ the global
optimum.
\end{proposition}

\begin{proof}
Pairing and precedence hold because $\mathcal{R}_k$ contains whole requests, so
$p_r$ and $d_r$ lie in the same subproblem and the route orders them. Capacity and
timing are subproblem constraints. Each request lies in exactly one
$\mathcal{R}_k$, so each is served once. The objective is a sum over couriers of
travel, a sum over stops of penalties, and a sum over couriers of fixed cost; all
three decompose along the partition, the last because $u_k$ is determined by
whether $\mathcal{R}_k$ is empty. Since the partition is one feasible choice among
many, the assembled cost is an upper bound on the global optimum.
\end{proof}

\subsection{Guarantees and their limits}\label{sec:guarantees}
Two of the five steps terminate provably. The descent of
Section~\ref{sec:descent} strictly decreases $\Phi$ of~\eqref{eq:descentobj}
over a finite set of assignments, so it halts at a local minimum of $\Phi$ with
respect to relocate and swap; and the assignment
program~\eqref{eq:assignprog} is solved to proven optimality for the
costs it is given.

The outer loop has no such guarantee. Each pass recomputes $C_{rk}$ from the current assignment, so
successive passes minimise \emph{different} linear objectives. In other words, there is no
single quantity the iteration descends, and the true cost may rise from one pass
to the next. What our proposed method relies on instead is bookkeeping. Every pass ends by routing its
assignment with the exact router of Section~\ref{sec:routing}, and the loop
returns the cheapest assignment it has seen rather than the one the last pass
happened to produce. Running further passes can therefore never worsen the
answer, whatever the passes themselves do.

Why $C_{rk}$ has to be recomputed at every pass is worth making precise, since
the same argument explains why $Q_k$ is kept small for a reason beyond
tractability.
Write $J(\mathcal{S},k)$ for the exact cost of serving a set of requests
$\mathcal{S}$ on courier $k$. If $J(\cdot,k)$ were \emph{modular}, that is
$J(\mathcal{S},k)=\sum_{r\in\mathcal{S}} j(r,k)$, then $C_{rk}=j(r,k)$ wherever
it was measured, the objective~\eqref{eq:assign} would equal the true cost,
and one pass would solve the assignment exactly. Routing is not modular: two
requests placed together interact. Write
\begin{equation}\label{eq:interaction}
I(\mathcal{S};r,r')=J(\mathcal{S}\cup\{r,r'\},k)-J(\mathcal{S}\cup\{r\},k)
-J(\mathcal{S}\cup\{r'\},k)+J(\mathcal{S},k)
\end{equation}
for the interaction of requests $r$ and $r'$ against a base set of requests
$\mathcal{S}$. By Proposition~\ref{prop:decompvalid} the
objective~\eqref{eq:globalobj} is $\sum_k J(\mathcal{A}_k,k)$ plus $\varphi$
times the couriers opened, and the program's
objective~\eqref{eq:assign} replaces each $J(\mathcal{A}_k,k)$ by
$\sum_{r\in\mathcal{A}_k}C_{rk}$ while pricing the fleet term exactly, since
that term depends only on which couriers are non-empty. The whole discrepancy
therefore sits in the per-courier terms. Under
modularity the two terms~\eqref{eq:interaction} adds and the two it subtracts
come to the same number, both being
$2J(\mathcal{S},k)+j(r,k)+j(r',k)$, so $I$ is identically zero; and $I$
identically zero in turn forces $J(\cdot,k)$ to be modular. The interaction is
therefore an exact measure of the additivity that~\eqref{eq:assign} assumes and
routing does not have.

Recomputing $C_{rk}$ at the start of every pass is what stops that error
compounding. The marginal costs are fixed for the length of a pass, while the assignment
program and the descent move requests between couriers: measured on the
instances of Section~\ref{sec:setup}, about a quarter of all requests change
courier within a single pass. By the end of one, many marginals therefore
describe workloads their couriers no longer hold, and the interaction
of~\eqref{eq:interaction} is what that drift costs. Measuring again at the assignment the pass actually
reached throws those stale marginals away instead of carrying them into the
next pass, so the error the program works under is the error of one pass rather
than of every pass so far. Whatever error survives both, it does not reach the
costs this paper reports. Every iterate is routed exactly by the label setting
of Section~\ref{sec:routing}, and the solution returned is the cheapest of those
exactly routed iterates, so each reported cost is the true cost of the solution
it describes, whatever the surrogate said while that solution was being built.
The surrogate can therefore steer the program to a worse assignment than exact
marginals would have chosen; it cannot make the cost reported for the assignment
that is returned look any better than that assignment really is.

\subsection{Parallel structure, and the cost of the coupled
step}\label{sec:parallel}\label{sec:lagrangian}
Four of the five steps do their expensive work per courier: the candidate
lists, the marginal costs of~\eqref{eq:marginal}, the screening and scoring
evaluations of the descent, and the exact routing of
Section~\ref{sec:routing}. What parallelises is that evaluation, not the step
around it. Building $\mathcal{C}(r)$ needs every courier's position broadcast
before the per-request lists can form; a relocate or swap changes two couriers,
so its $\Delta\Phi$ assembles from two local evaluations; and the descent
applies the best move of a sweep, a reduction over all of them. Between those
synchronisation points the per-courier evaluations run independently; the sweep
as a whole does not. Only the exact routing is independent end to end, since the
partition is fixed before it starts and the subproblems share nothing. The
assignment program~\eqref{eq:assignprog} does not decompose at all, because the
covering constraint~\eqref{eq:cover} ties the couriers into one program.
Section~\ref{sec:scaling} reports the measured split.

What follows asks what it would take to remove that one exception, and what
removing it costs. The way to remove it is to charge for the covering constraint
rather than impose it. Write
$\mathcal{L}_k=\{r\in\mathcal{R}:k\in\mathcal{C}(r)\}$ for the requests whose
candidate list contains courier $k$, the index set of~\eqref{eq:cap}, and
$x_{\cdot k}=(x_{rk})_{r\in\mathcal{L}_k}$ for the assignment variables of
courier $k$ alone. Attaching a
multiplier $\lambda_r$ to each covering constraint~\eqref{eq:cover} gives
\begin{equation}\label{eq:lagr}
L(\lambda)=\sum_r \lambda_r+\sum_{k\in\mathcal{K}}v_k(\lambda),
\qquad
v_k(\lambda)=\min_{x_{\cdot k},u_k}
\Bigl[\sum_{r\in\mathcal{L}_k} (C_{rk}-\lambda_r)x_{rk}+\varphi u_k\Bigr],
\end{equation}
where $v_k(\lambda)$ is minimised subject to constraint~\eqref{eq:cap} for
courier $k$ and integrality, and to nothing else. Two properties
of~\eqref{eq:lagr} carry the parallel step and are worth stating exactly: the
$v_k(\lambda)$ share no variable, and each is evaluated by a selection rather
than by a solver.

\begin{proposition}[The priced assignment separates]\label{prop:lagr}
Fix multipliers $\lambda\in\mathbb{R}^{|\mathcal{R}|}$ and write
$\bar{C}_{rk}=C_{rk}-\lambda_r$ for reduced costs. Then $L(\lambda)$
of~\eqref{eq:lagr} is the optimal value
of~\eqref{eq:assignprog} with the covering
constraint~\eqref{eq:cover} deleted from the feasible set and priced in the
objective instead. In particular $L(\lambda)$ is at most the optimal value
of~\eqref{eq:assignprog}, for every
$\lambda$. No variable appears in both $v_k(\lambda)$ and $v_{k'}(\lambda)$ for
$k\ne k'$. Finally, let
$\mathcal{G}_k=\{r\in\mathcal{L}_k:\bar{C}_{rk}<0\}$ be the requests courier $k$
would gain by taking, and let $\mathcal{S}_k\subseteq\mathcal{G}_k$ consist of
the $\min\{Q_k,|\mathcal{G}_k|\}$ members of $\mathcal{G}_k$ of smallest reduced
cost, ties broken arbitrarily, so that $\mathcal{S}_k=\emptyset$ when
$\mathcal{G}_k=\emptyset$. Writing
$\sigma_k=\sum_{r\in\mathcal{S}_k}\bar{C}_{rk}$, with $\sigma_k=0$ for an empty
sum, $v_k(\lambda)=\min\{0,\;\varphi+\sigma_k\}$: when $\sigma_k<-\varphi$ the
minimiser is $u_k=1$ with $x_{\cdot k}$ the indicator vector of
$\mathcal{S}_k$, and otherwise it is $u_k=0$, $x_{\cdot k}=0$.
\end{proposition}

\begin{proof}
\emph{The dual function.} Constraint~\eqref{eq:cover} is an equality, so its
multiplier is free in sign. Adding
$\sum_r\lambda_r(1-\sum_{k\in\mathcal{C}(r)}x_{rk})$ to~\eqref{eq:assign} and
collecting the terms that carry a variable,
\[
\sum_{r}\sum_{k\in\mathcal{C}(r)}C_{rk}x_{rk}+\varphi\sum_{k}u_k
+\sum_r\lambda_r\Bigl(1-\sum_{k\in\mathcal{C}(r)}x_{rk}\Bigr)
=\sum_r\lambda_r+\sum_{r}\sum_{k\in\mathcal{C}(r)}\bar{C}_{rk}x_{rk}
+\varphi\sum_{k}u_k ,
\]
since each $x_{rk}$ is charged $C_{rk}$ once and credited $\lambda_r$ once.
Minimising over the constraints that remain, which are~\eqref{eq:cap} and
integrality, and taking the constant $\sum_r\lambda_r$ outside the minimisation
gives~\eqref{eq:lagr}. Any $(x,u)$ feasible
for~\eqref{eq:assignprog} satisfies~\eqref{eq:cover}, so the added
term vanishes at it and its objective value is unchanged, while the minimisation
also ranges over points that violate~\eqref{eq:cover}; hence $L(\lambda)$ is at
most the optimal value of~\eqref{eq:assignprog}.

\emph{Separability.} The surviving constraints are constraint~\eqref{eq:cap},
one instance of it per courier, together with integrality; the variables
$(x_{\cdot k},u_k)$ appearing in courier $k$'s instance appear in no other
courier's. The feasible set is therefore the product $\prod_{k}X_k$ with
$X_k=\{(x_{\cdot k},u_k)\in\{0,1\}^{|\mathcal{L}_k|+1}:
\sum_{r\in\mathcal{L}_k}x_{rk}\le Q_k u_k\}$. Exchanging the order of
summation,
$\sum_{r\in\mathcal{R}}\sum_{k\in\mathcal{C}(r)}\bar{C}_{rk}x_{rk}
=\sum_{k\in\mathcal{K}}\sum_{r\in\mathcal{L}_k}\bar{C}_{rk}x_{rk}$, both sides
running over the same pairs, so the objective is a sum whose $k$th term depends
on the $k$th factor alone. A minimum of such a sum over a product set is the sum
of the minima of its terms, which is $\sum_k v_k(\lambda)$.

\emph{Each courier's minimum.} Fix $k$. If $u_k=0$ then~\eqref{eq:cap} forces
$x_{\cdot k}=0$, at an objective value of $0$. If $u_k=1$ the objective is
$\varphi+\min\{\sum_{r\in\mathcal{S}}\bar{C}_{rk}:
\mathcal{S}\subseteq\mathcal{L}_k,\;|\mathcal{S}|\le Q_k\}$, and the inner
minimum is attained at $\mathcal{S}_k$. Removing from $\mathcal{S}$ a request
with $\bar{C}_{rk}\ge0$ does not increase the sum and cannot violate capacity,
so an optimal $\mathcal{S}$ may be taken inside $\mathcal{G}_k$; if
$r\in\mathcal{S}$ and $r'\in\mathcal{G}_k\setminus\mathcal{S}$ satisfy
$\bar{C}_{r'k}<\bar{C}_{rk}$ then exchanging them strictly decreases the sum, so
the requests held are those of smallest reduced cost in $\mathcal{G}_k$; and
whenever $|\mathcal{S}|<Q_k$ and some member of $\mathcal{G}_k$ is left,
adding it decreases the sum further. The selection therefore stops only on
running out of capacity or out of $\mathcal{G}_k$, whichever comes first, which
is $\mathcal{S}_k$. Comparing the two
cases for $u_k$ gives $v_k(\lambda)=\min\{0,\varphi+\sigma_k\}$, with the
minimiser as stated.
\end{proof}

\noindent A courier therefore needs nothing from any other courier: its own
reduced costs, its own capacity and $\varphi$ settle both what it claims and
whether it opens, and they settle it by a selection among $|\mathcal{L}_k|$
numbers rather than by a call to a solver. That is what a machine with one core
per courier would run in parallel.

What this does \emph{not} do is enforce~\eqref{eq:cover}: two couriers may claim
the same request and none may claim another, so the claims that come with the
bound of Proposition~\ref{prop:lagr} are not an assignment. Prices are updated
by the subgradient step
$\lambda_r \leftarrow \lambda_r + \text{step}\cdot(1-\sum_k x_{rk})$, which
lowers $\lambda_r$ when more than one courier claimed $r$ and raises it when
none did. A courier sees $r$ only through the reduced cost $C_{rk}-\lambda_r$,
so a contested request becomes less attractive to all of them and an unclaimed
one more attractive, and a repair step turns the claims into a feasible
assignment.

Because~\eqref{eq:assignprog} is an integer program, $L(\lambda)$ is a
lower bound that need not be tight, and a duality gap may persist even at the
best multipliers. Section~\ref{sec:lagrangianresults} reports not that gap but
the \emph{solution loss}: what the priced step and its repair cost against the
exact program on the same instance. That is the quantity a practitioner pays.
The bound itself we do not report, because the proposed method solves the
assignment exactly and never forms $L(\lambda)$. The loss is
concentrated in the fixed charge: the multipliers coordinate \emph{which} courier takes a request well,
and \emph{whether a courier opens at all} badly. We therefore use the exact
assignment in the proposed method and report the Lagrangian variant as the
option to take when the assignment must itself be distributed.

%%%%%%%%%%%%%%%%%%%%%%%%%%%%%%%%%%%%%%%%
%%%%      THE ROUTING SUBPROBLEM    %%%%
%%%%%%%%%%%%%%%%%%%%%%%%%%%%%%%%%%%%%%%%

%%%%%%%%%%%%%%%%%%%%%%%%%%%%%%%%%%%%%%%%
%%%%           EXPERIMENTS          %%%%
%%%%%%%%%%%%%%%%%%%%%%%%%%%%%%%%%%%%%%%%

\section{Experiments}\label{sec:sims}

\subsection{Experimental setup}\label{sec:setup}
\titledParagraph{Instances.} Instances are generated from a rule that makes the
fleet decision non-trivial by construction. Given the fleet size $|\mathcal{K}|$, a
reduction factor $f$ (the share of the fleet a good solution is expected to
dispatch) and $\nu$ (requests per dispatched courier),
\[
K_{\mathrm{eff}}=\operatorname{round}(f|\mathcal{K}|),\quad
|\mathcal{R}|=\nu K_{\mathrm{eff}},\quad
Q_k=\lfloor 4\nu/3 \rfloor ,
\]
which forces $\lceil|\mathcal{R}|/Q_k\rceil$ couriers at minimum and permits all
$|\mathcal{K}|$, so how many are dispatched is left to $\varphi$. We use $f=5/6$ and $\nu=4$ throughout,
which fixes $Q_k = 5$ and leaves the fleet free between $\lceil
|\mathcal{R}|/5\rceil$ and $|\mathcal{K}|$. The main comparison uses the three
sizes this rule gives at $|\mathcal{K}| = 12$, $30$ and $60$, namely
$|\mathcal{R}| = 40$, $100$ and $200$; at $|\mathcal{K}| = 12$, for instance, a
solution may dispatch anywhere between $8$ and $12$ couriers. Two alignment
settings and three random seeds at each size give 18 instances.

\begin{figure}[!htbp]\centering
\includegraphics[width=\textwidth]{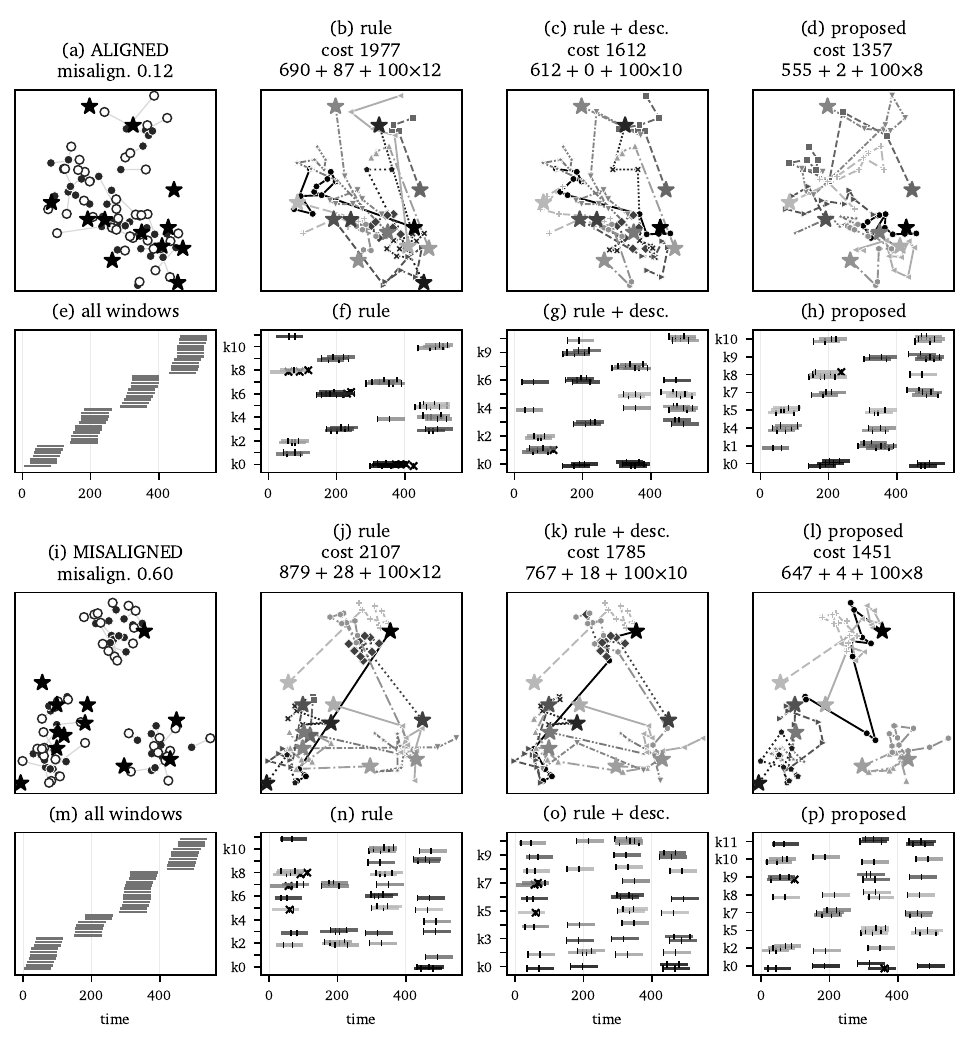}
\caption{Both regimes on the same axes, $|\mathcal{R}|=40$ requests,
$|\mathcal{K}|=12$ couriers available, $Q_k=5$, $\varphi=100$. Rows~1--2 are an aligned instance
(misalignment $0.12$: requests that are close in space are also close in time);
rows~3--4 a misaligned one (misalignment $0.60$). Within each pair, the map row
shows the instance as posed, with filled circles for pickups, open circles
for deliveries and stars for the twelve couriers available; then the routes found
by capacitated $k$-means as published, labelled \emph{rule}, by the same rule
carrying our descent, and by the proposed method; the row beneath shows every
request's window, then the windows of each dispatched courier's own requests;
the horizontal axis of every panel in those rows is time, labelled once at the
foot of the figure. A $|$ marks when the exact route serves a stop, $\times$ a late
one. Below each map panel's cost are its three terms in the order
of~\eqref{eq:globalobj}: travel, lateness, and $\varphi$ times the couriers
dispatched. Cost falls $1977 \rightarrow 1612 \rightarrow 1357$ when aligned and
$2107 \rightarrow 1785 \rightarrow 1451$ when misaligned: travel, tardiness and
fleet all improve in both. The published rule dispatches all twelve couriers in
either regime because it has no mechanism to do otherwise, while the proposed
method dispatches eight, which is why eight rows of the timeline are occupied
rather than twelve. Each instance shown is the median of
five seeds, not the best.}
\label{fig:mechanism}
\end{figure}

\begin{figure}[!htbp]\centering
\includegraphics[width=\textwidth]{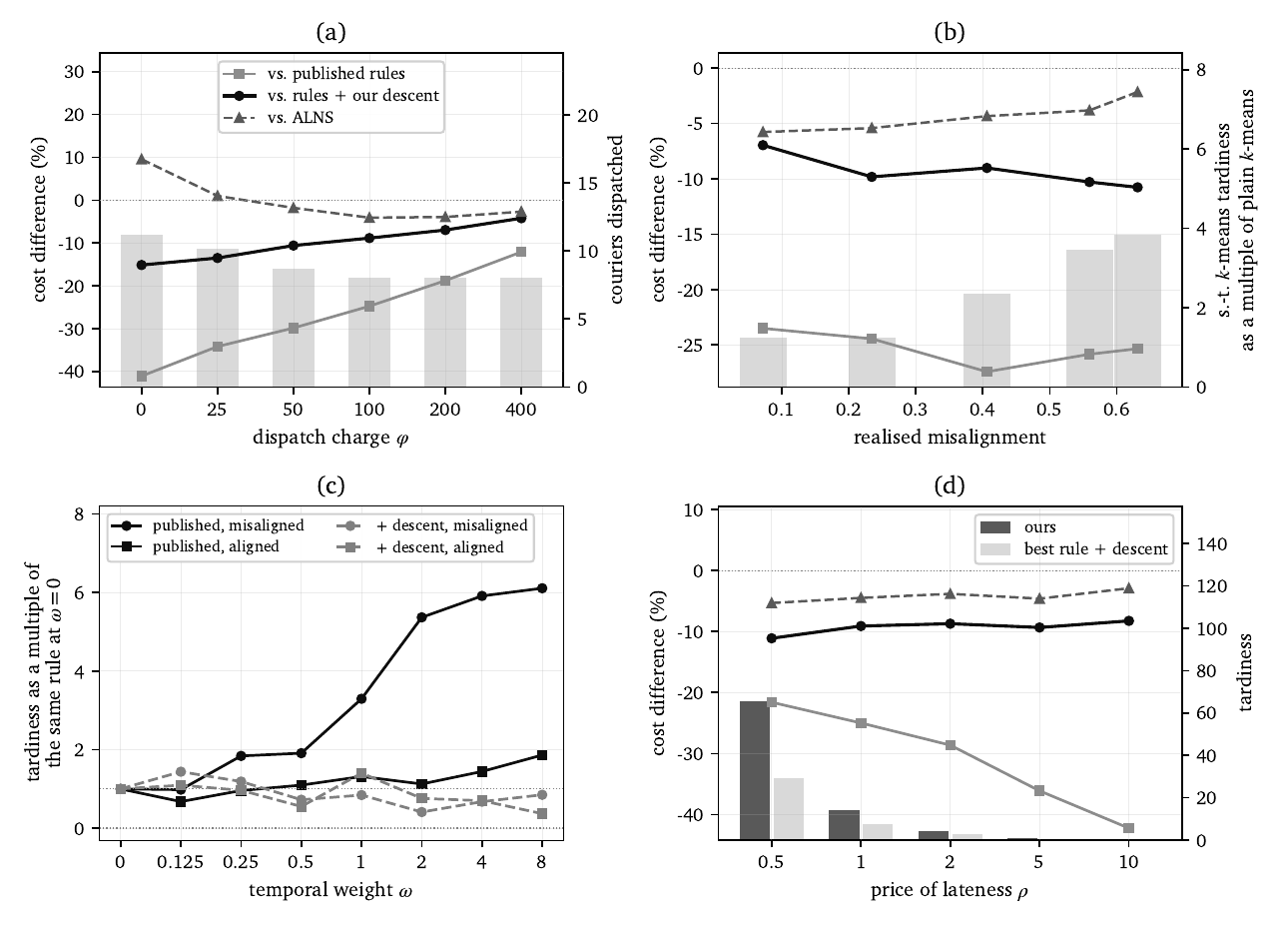}
\caption{The four parameter sweeps. In every panel the lines are cost
differences against the baselines on the left axis, negative favouring the
proposed method, and the bars on the right axis carry the quantity that
explains the trend. (a) the dispatch charge, with the fleet we open; (b) the alignment
axis, with the tardiness spatio-temporal $k$-means pays for its time coordinate
relative to plain $k$-means; (c) the temporal weight $\omega$ of
Section~\ref{sec:setup}, with tardiness as a multiple of the same rule at
$\omega=0$; (d) the penalty rate $\rho$, with the
tardiness of both methods. Instances are $|\mathcal{R}|=40$,
$|\mathcal{K}|=12$, $Q_k=5$, with $\varphi=100$ and $\rho=1$ except on the axis
being swept.}
\label{fig:sensitivity}
\end{figure}

Within that frame an instance is drawn as follows. Four spatial hubs are placed
uniformly in a square of side $100$, and each request is assigned a hub; its
pickup is Gaussian about the hub with standard deviation $5$, and its delivery
Gaussian about its own pickup with the same spread, so a request is local to its
hub. Travel times are Euclidean distances at unit speed, so a distance and a
duration are the same number. The horizon, from time zero to the last delivery deadline, is divided into bands
separated by a fixed gap. \emph{Each
request's band is copied from its spatial hub with probability $\alpha$, and drawn
independently of it otherwise}: $\alpha=1$ makes spatial neighbours temporal
neighbours, $\alpha=0$ makes the two structures unrelated, and the realised
misalignment is measured rather than assumed. \emph{Misalignment} is one minus
the agreement between two four-way Ward partitions of the requests, one on their
spatial midpoints and one on the midpoints of their intervals, maximised over the
ways of matching their parts: $0$ means the spatial and temporal
groupings coincide, and a value near $1$ means they are unrelated. Pickup windows are $45$ wide and a
request's delivery window has the same width shifted $35$ later, so the
\emph{request interval}, the span from a request's earliest pickup to its latest
delivery, is $45+35=80$ time units for every request. Service takes $8$ at each
stop. Couriers start at perturbed hub
positions and become available at staggered times over the first $5\%$ of the
horizon. We report $\alpha\in\{1,0\}$.
\figref{fig:mechanism} shows one instance at each of those two
settings, with three solutions drawn side by side on both: capacitated
$k$-means as published, the same rule carrying our descent, and the proposed
method. Four of the settings fixed here are swept one at a time in the
subsections that follow, and \figref{fig:sensitivity} collects those sweeps in
one place: the dispatch charge in panel~(a), the alignment of space and time
in~(b), the weight on a temporal coordinate in~(c), and the penalty rate
in~(d).

\titledParagraph{What this design does and does not cover.} Holding the window
width and the delivery shift constant is what isolates the alignment
axis, but it also makes the temporal scale uniform: every instance has the same
request interval, and the fraction of request pairs whose intervals overlap sits
near $0.25$. Because each delivery is drawn about its own pickup, the
\emph{locality} of these instances, the fraction of requests whose pickup and
delivery lie within $10$ of each other, is high;
Section~\ref{sec:instancespace} lowers it deliberately by sending deliveries to
a different hub. The \emph{courier-time ratio} is $|\mathcal{K}|$ times the span from
the earliest window opening or courier ready time to the last deadline, divided
by the total service time at all stops plus $2|\mathcal{R}|$ times the median
distance between two stops of the instance. Below $1$ the fleet has less time than the work is estimated to
need, so no grouping of the requests can be punctual and the instance measures
which method pays the smallest forced penalty. Adapted Li \& Lim batches at the same
$|\mathcal{R}|$, $|\mathcal{K}|$
and $Q_k$ differ in all three respects: mean interval $465$ against
our $80$, overlap $0.77$ against $0.25$, and courier-time ratios from $0.57$ to
$3.95$ against our $1.6$--$1.8$. They also occupy only the misaligned end of the
axis, with measured misalignment $0.55$--$0.68$ against our $0.00$--$0.70$.
Section~\ref{sec:lilim} therefore repeats the comparison on them.

\titledParagraph{Adapting the benchmark batches.} The Li \& Lim benchmark is
specified for the classical PDPTW, which sends every vehicle out from one depot
and back to it, treats time windows as hard, and measures demand by weight.
Section~\ref{sec:problemFormulation} assumes none of those three, so each
instance is converted before use, as follows.
A dispatch batch of $|\mathcal{R}|$ requests is sampled from the
requests released inside one window of the benchmark day, since a batch drawn
across a whole day is one no fleet can serve. The benchmark's demands are
discarded and every request counts as one unit against $Q_k$. Routes are open
and windows are soft. Couriers start at perturbed depot positions and become
ready at staggered times, as in the generated instances. The $\texttt{lc1}$
family is excluded: its service times exceed its window widths, so no
multi-stop route can be punctual and every method is dominated by forced
lateness. The fleet, capacity and request count are set to the same design
point, $|\mathcal{R}|=40$, $|\mathcal{K}|=12$, $Q_k=5$, so the two tables may be
read against each other.

\titledParagraph{Baselines.} We compare the proposed method against six
clustering rules from the literature, the groupings a cluster-first dispatcher
would otherwise use. They are \textsc{sweep}
\citep{gillett1974heuristic}, \textsc{fisher}~\citep{fisher1981generalized},
$k$-means~\citep{lloyd1982least} on request midpoints, spatio-temporal
$k$-means, which is the same algorithm on a third coordinate
$\omega\,(\sigma_x/\sigma_\tau)\,\tau_r$ alongside the midpoint, where $\tau_r$
is the temporal midpoint of Section~\ref{sec:cost}, $\sigma_x$ and
$\sigma_\tau$ are the spatial and temporal spreads of the instance, and the
dimensionless \emph{temporal weight} $\omega$ sets how far the time axis is
stretched relative to the spatial ones, so that $\omega=0$ recovers plain
$k$-means and $\omega=1$ gives time the same spread as space, capacitated
$k$-means, which assigns requests to centroids by the transportation problem
of~\citet{bradley2000constrained} with an upper bound of $Q_k$ requests per
cluster in place of their lower bound, and Ward agglomerative
clustering~\citep{ward1963hierarchical}. Two values are reported for each method and they answer different questions. \emph{As published} measures our proposed
method against what the literature proposes. \emph{Strengthened} attaches our descent of
Section~\ref{sec:descent} to the rule's own grouping, so that both sides improve
their assignment with the same step. That also gives the rule a response to
$\varphi$, which no bare rule has, since the descent minimises the objective
and the objective contains $\varphi$. This second
comparison therefore answers two questions: how much a generic local search
recovers for an existing rule, and, with the improvement step now common to both
sides, how much is gained by grouping on the marginal cost
of~\eqref{eq:marginal} rather than on the geometric criteria these rules use.
All six rules are routed with the exact router of Section~\ref{sec:routing}, the
one our own method uses, so none of the difference between them and us comes from
route quality.
% The reference of the next paragraph is the exception: it searches sequences itself, so
% its routes are its own, and every cost it reports is re-evaluated with our
% scheduler to keep the objective comparable.

\titledParagraph{ALNS.} We also compare against an adaptive large neighbourhood
search~\citep{ropke2006adaptive}, which produced many of the best known solutions
for the standard PDPTW benchmarks and remains the usual reference point for the
problem. It is not a cluster-first method and forms no groups at all, which makes
it the opposite kind of algorithm to the six above and to ours. It holds a
complete solution, and each iteration removes part of that solution and reinserts
the removed requests wherever they fit best, so which courier serves a request
and where in that courier's sequence it falls are settled together and revised
continually, never fixed in stages. The rules deciding what to remove and where to put it back
are the standard ones and we changed none of them. All are in play: removal at
random, by the saving it yields, and by relatedness between requests in the sense
of \citet{shaw1998using}; reinsertion at the cheapest position and by
regret~\citep{ropke2006adaptive}. What the search adapts is the choice among
them: a roulette wheel reweights them towards those that have recently improved
the solution, and a simulated-annealing rule decides whether an iteration's
result is kept.

Every one of those rules works over the whole solution, and that
is what makes ALNS a \emph{centralized} method. We use that word
literally rather than as a label: a removal rule ranks every request in
the instance against the objective of the whole current solution, and reinsertion
tries every position in every route, so no step of it can be
carried out knowing one courier's requests alone. That is the property the
decomposition gives up, and Section~\ref{sec:parallel} states how much of ours
retains it: the expensive work of four of the five steps is computed from
one courier's data, and only the assignment program~\eqref{eq:assignprog} is
irreducibly joint. We therefore include
ALNS as a reference point rather than a competitor, and hold it to the same
conditions. Its budget is \emph{the wall-clock our proposed method used on that
instance}, subject to a floor of ten seconds so that the smallest instances are
not searched for a second and then stopped. The floor binds wherever we finish
in under ten seconds, which is every instance at
$|\mathcal{R}| = 40$, where we take about one second and ALNS is therefore given
ten; the comparison favours ALNS by that much and we report it as run. The
budget can only be tested between iterations, so the search finishes the
ruin-and-recreate pass it is in when the budget expires, which at these sizes
costs a fraction of a second and at the sizes of
Section~\ref{sec:scaling} costs a great deal more. ALNS obeys the total-count
capacity of Section~\ref{sec:data}, and, being stochastic, is run three times
with the median reported. It is also the one method that sequences for itself, so its routes are
its own rather than our router's; every solution it returns is re-costed under
the earliest-feasible schedule of Remark~\ref{rem:earliness}, so that the
objective compared is the same one.

\titledParagraph{Protocol.} All comparisons are paired on identical instances
and reported with win counts and a Wilcoxon signed-rank $p$-value. Travel,
tardiness and the courier count are reported separately as well as in the
objective~\eqref{eq:globalobj}, since a method can win on one and lose on
another. Our proposed method is run once on each instance. Its only use of a
random number is a perturbation of at most $10^{-6}$ that breaks ties when the
construction orders requests by release time, and no two release times in these
instances are that close, so a restart from a different seed returns the same
answer. ALNS is stochastic and is run three times with the median reported:
restarted five times on twelve instances it spread $11.5\%$ in cost on average,
so a single run of it would not measure what it can do.
Machine: Intel i7-12800H, Python 3.12, HiGHS via SciPy.

% ===== BEGIN GENERATED RESULTS =====
% Everything between these markers is written by code/make_tables.py.
% Re-run it after any experiment; hand edits here are overwritten.
\subsection{Comparison with published clustering rules}\label{sec:rules}

\tabref{tab:main} reports every method on 18 instances across
3 problem sizes, at $\varphi = 100$, two alignment
levels and three seeds each. Its cost columns are means over the 18
instances, and its \emph{vs.\ ours} columns are paired on the per-instance
differences instead, since the instances differ in difficulty by more than the
methods differ from one another.

\begin{table}[t]\centering
\sbox{\tabbox}{\small\begin{tabular}{@{}l rrrrr rrrrr@{}}
\toprule
& \multicolumn{5}{c}{as published} & \multicolumn{5}{c}{+ descent}\\
\cmidrule(lr){2-6}\cmidrule(lr){7-11}
method & travel & tard. & cour. & cost & vs.\ ours & travel & tard. & cour. & cost & vs.\ ours\\
\midrule
\textsc{sweep} & 2163 & 833 & 22.7 & 5263 & $-29.3\%$ & 1932 & 35 & 22.7 & 4234 & $-13.3\%$\\
\textsc{fisher} & 2487 & 578 & 34.0 & 6464 & $-42.6\%$ & 1692 & 12 & 26.6 & 4359 & $-16.1\%$\\
$k$-means & 2174 & 410 & 34.0 & 5984 & $-38.8\%$ & 1876 & 18 & 28.1 & 4700 & $-21.6\%$\\
spatio-temporal $k$-means & 2217 & 800 & 34.0 & 6417 & $-43.0\%$ & 1932 & 31 & 28.6 & 4819 & $-22.7\%$\\
capacitated $k$-means & 2284 & 449 & 34.0 & 6133 & $-40.0\%$ & 1922 & 12 & 28.4 & 4773 & $-22.5\%$\\
Ward agglomerative & 2176 & 326 & 34.0 & 5902 & $-36.8\%$ & 1822 & 40 & 28.2 & 4684 & $-19.5\%$\\
\addlinespace[2pt]
best of six & 2149 & 809 & 22.9 & 5247 & $-28.8\%$ & 1779 & 18 & 23.9 & 4186 & $-12.1\%$\\
\midrule
\textbf{this paper} & \multicolumn{5}{c}{} & 1206 & 105 & 23.2 & \textbf{3628} & ---\\
ALNS & \multicolumn{5}{c}{} & 2040 & 92 & 23.1 & 4438 & $-14.2\%$\\
\bottomrule\end{tabular}}%
\setlength{\tabw}{1.00\textwidth}%
\ifdim\wd\tabbox>\tabw\setlength{\tabw}{\wd\tabbox}\fi
\begin{minipage}{\tabw}
\centering
\caption{$|\mathcal{R}| = 40, 100, 200$ requests, $|\mathcal{K}| = 12, 30, 60$ couriers, $Q_k = 5$, $\varphi = 100$, $\rho = 1$. Cost and its components, means over all instances. \emph{As published} uses the rule in its original form; \emph{+ descent} adds the improvement step of Section~\ref{sec:descent}. \emph{vs.\ ours} is the mean paired percentage cost difference against the proposed method, negative favouring it; each is a Wilcoxon signed-rank test over the same instances, and every one of them has $p<10^{-3}$. \emph{Best of six} takes the cheapest rule on each instance separately, so it is a stricter baseline than any single row. The proposed method and ALNS have no \emph{as published} form, so they appear once. Cost is travel${}+{}$tardiness${}+\varphi\times{}$couriers.}
\label{tab:main}
\small\begin{tabular*}{\tabw}{@{\extracolsep{\fill}}l rrrrr rrrrr@{}}
\toprule
& \multicolumn{5}{c}{as published} & \multicolumn{5}{c}{+ descent}\\
\cmidrule(lr){2-6}\cmidrule(lr){7-11}
method & travel & tard. & cour. & cost & vs.\ ours & travel & tard. & cour. & cost & vs.\ ours\\
\midrule
\textsc{sweep} & 2163 & 833 & 22.7 & 5263 & $-29.3\%$ & 1932 & 35 & 22.7 & 4234 & $-13.3\%$\\
\textsc{fisher} & 2487 & 578 & 34.0 & 6464 & $-42.6\%$ & 1692 & 12 & 26.6 & 4359 & $-16.1\%$\\
$k$-means & 2174 & 410 & 34.0 & 5984 & $-38.8\%$ & 1876 & 18 & 28.1 & 4700 & $-21.6\%$\\
spatio-temporal $k$-means & 2217 & 800 & 34.0 & 6417 & $-43.0\%$ & 1932 & 31 & 28.6 & 4819 & $-22.7\%$\\
capacitated $k$-means & 2284 & 449 & 34.0 & 6133 & $-40.0\%$ & 1922 & 12 & 28.4 & 4773 & $-22.5\%$\\
Ward agglomerative & 2176 & 326 & 34.0 & 5902 & $-36.8\%$ & 1822 & 40 & 28.2 & 4684 & $-19.5\%$\\
\addlinespace[2pt]
best of six & 2149 & 809 & 22.9 & 5247 & $-28.8\%$ & 1779 & 18 & 23.9 & 4186 & $-12.1\%$\\
\midrule
\textbf{this paper} & \multicolumn{5}{c}{} & 1206 & 105 & 23.2 & \textbf{3628} & ---\\
ALNS & \multicolumn{5}{c}{} & 2040 & 92 & 23.1 & 4438 & $-14.2\%$\\
\bottomrule\end{tabular*}
\end{minipage}
\end{table}

The \emph{vs.\ ours} columns of \tabref{tab:main} pair the proposed method
against each of the six rules separately and against the best of them. The
comparison we lead with is the last, and it is the strictest: on each instance
the baseline is whichever of the six is cheapest \emph{on that instance}, so
the margin against it is necessarily smaller than the margin against any rule
taken alone, as the spread of those rows shows. Against that baseline our proposed method is
29\% cheaper with the six as published, winning 18
of 18 instances. Giving the six our own descent narrows that to
12.1\% ($p<10^{-3}$, still 18/18). The two
figures measure different things. The
16.7 points between them are what a generic
improvement step recovers for an existing rule. The
12.1\% that survives it is the contribution of
grouping on the marginal cost of~\eqref{eq:marginal}, measured with the
improvement step common to both sides, and it is the figure the rest of this
section builds on.

The cost difference behind that 12.1\% is the sum of three
terms of the objective~\eqref{eq:globalobj} that move in different directions.
Pooled over the 18 instances and measured in the units of that objective,
our proposed method saves $573$ in travel, gives up $87$ in
tardiness and opens $0.72$ fewer couriers on average,
worth $72$ at $\varphi=100$; the three account exactly
for the $558$ it saves in cost. The advantage is therefore a
large reduction in travel bought with a small concession in lateness, and not a
uniform improvement across the objective. The margin also grows with instance size, and
\tabref{tab:bysize} gives it by size against both baselines. Against the six
strengthened rules it is $-8.9\%$ at $|\mathcal{R}|=40$, $-13.1\%$ at $|\mathcal{R}|=100$, $-14.4\%$ at $|\mathcal{R}|=200$.

\begin{table}[t]\centering
\sbox{\tabbox}{\small\begin{tabular}{@{}r rrr rrr@{}}\toprule
& \multicolumn{3}{c}{vs.\ best of six $+$ descent} & \multicolumn{3}{c}{vs.\ ALNS}\\
\cmidrule(lr){2-4}\cmidrule(lr){5-7}
$|\mathcal{R}|$ & travel & cour. & cost & travel & cour. & cost\\\midrule
$40$ & $-9.3\%$ & $-8.9\%$ & $-8.9\%$ & $-3.6\%$ & $-3.7\%$ & $-4.1\%$\\
$100$ & $-32.9\%$ & $+1.1\%$ & $-13.1\%$ & $-37.4\%$ & $+0.9\%$ & $-17.6\%$\\
$200$ & $-36.4\%$ & $-3.0\%$ & $-14.4\%$ & $-47.1\%$ & $+1.3\%$ & $-20.8\%$\\
\bottomrule\end{tabular}}%
\setlength{\tabw}{0.80\textwidth}%
\ifdim\wd\tabbox>\tabw\setlength{\tabw}{\wd\tabbox}\fi
\begin{minipage}{\tabw}
\centering
\caption{$|\mathcal{R}| = 40, 100, 200$ requests, $|\mathcal{K}| = 12, 30, 60$ couriers, $Q_k = 5$, $\varphi = 100$, $\rho = 1$. Paired differences by problem size, on the instances of \tabref{tab:main}. Each entry is the mean of the per-instance percentage differences at that size, negative favouring the proposed method. \emph{Best of six} is the cheapest of the six rules carrying our descent on that instance. Six instances at each size, two alignment levels by three seeds. A tardiness column is omitted because both baselines are close to punctual here, so a ratio against their tardiness is set by the instances where it is almost zero rather than by the difference; the pooled difference in the units of the objective is given in the text.}
\label{tab:bysize}
\small\begin{tabular*}{\tabw}{@{\extracolsep{\fill}}r rrr rrr@{}}\toprule
& \multicolumn{3}{c}{vs.\ best of six $+$ descent} & \multicolumn{3}{c}{vs.\ ALNS}\\
\cmidrule(lr){2-4}\cmidrule(lr){5-7}
$|\mathcal{R}|$ & travel & cour. & cost & travel & cour. & cost\\\midrule
$40$ & $-9.3\%$ & $-8.9\%$ & $-8.9\%$ & $-3.6\%$ & $-3.7\%$ & $-4.1\%$\\
$100$ & $-32.9\%$ & $+1.1\%$ & $-13.1\%$ & $-37.4\%$ & $+0.9\%$ & $-17.6\%$\\
$200$ & $-36.4\%$ & $-3.0\%$ & $-14.4\%$ & $-47.1\%$ & $+1.3\%$ & $-20.8\%$\\
\bottomrule\end{tabular*}
\end{minipage}
\end{table}
\subsection{The grouping cost examined}\label{sec:groupingcost}\titledParagraph{Non-additivity of the routing cost.}\label{sec:interaction}
Section~\ref{sec:guarantees} identifies the pairwise interaction of
$J(\cdot,k)$ as the exact measure of the additivity that the assignment program
assumes and routing does not have. This subsection measures it. We sampled 12,000 second differences
$I(\mathcal{S};r,r')=J(\mathcal{S}\cup\{r,r'\})-J(\mathcal{S}\cup\{r\})
-J(\mathcal{S}\cup\{r'\})+J(\mathcal{S})$ with the exact router, in which
$\mathcal{S}$ is a set of requests a courier already holds and $r,r'$ are two
further requests it might be given.
\tabref{tab:interaction} reports them by
$|\mathcal{S}|$, which runs from $0$ to $Q_k-2=3$: all of
$\mathcal{S}\cup\{r,r'\}$ has to fit one courier, so a base set of
4 leaves no room for a pair.

\begin{table}[t]\centering
\sbox{\tabbox}{\small\begin{tabular}{@{}r@{\;\;} r@{\;\;} r@{\;\;} r@{\;\;} r@{}}\toprule
$|\mathcal{S}|$ & samples & zero & median & 95th pct.\\\midrule
$0$ & 3,054 & $0\%$ & $0.538$ & $0.655$\\
$1$ & 3,068 & $27\%$ & $0.201$ & $0.462$\\
$2$ & 2,944 & $39\%$ & $0.074$ & $0.344$\\
$3$ & 2,934 & $45\%$ & $0.024$ & $0.261$\\
\bottomrule\end{tabular}}%
\setlength{\tabw}{0.80\textwidth}%
\ifdim\wd\tabbox>\tabw\setlength{\tabw}{\wd\tabbox}\fi
\begin{minipage}{\tabw}
\centering
\caption{$|\mathcal{R}| = 40$ requests, $|\mathcal{K}| = 12$ couriers, $Q_k = 5$, $\varphi = 100$, $\rho = 1$. The interaction $I(\mathcal{S};r,r')$ of~\eqref{eq:interaction}, computed with the exact router on sampled triples $(\mathcal{S},r,r')$ and grouped by $|\mathcal{S}|$, the number of requests the courier already holds. $I$ is zero for every $\mathcal{S}$, $r$ and $r'$ exactly when a courier's routing cost is additive across its requests, which is what the assignment program assumes. \emph{Zero} is the fraction of samples whose $I$ vanishes exactly; \emph{median} and \emph{95th pct.} give $|I|$ relative to the marginal it perturbs.}
\label{tab:interaction}
\small\begin{tabular*}{\tabw}{@{\extracolsep{\fill}}r@{\;\;} r@{\;\;} r@{\;\;} r@{\;\;} r@{}}\toprule
$|\mathcal{S}|$ & samples & zero & median & 95th pct.\\\midrule
$0$ & 3,054 & $0\%$ & $0.538$ & $0.655$\\
$1$ & 3,068 & $27\%$ & $0.201$ & $0.462$\\
$2$ & 2,944 & $39\%$ & $0.074$ & $0.344$\\
$3$ & 2,934 & $45\%$ & $0.024$ & $0.261$\\
\bottomrule\end{tabular*}
\end{minipage}
\end{table}

The routing cost is not additive across requests, and the table says by how
much. The typical departure from additivity is small, median
7.3 time units against a mean of 171 for
serving one of these sets, and its sign is close to balanced,
35\% negative against 38\% positive, so the errors a courier
accumulates over its requests partly cancel rather than compound. The mean
relative interaction is $-0.9\%$. In
27\% of samples it vanishes exactly, which happens when the two requests
never compete for a position in the route: each is inserted where the other
leaves the schedule untouched, so what one adds does not depend on whether the
other is there. The worst case is different: the largest interaction sampled is
307 time units, so Section~\ref{sec:guarantees} claims nothing about how
accurately any one courier's requests are costed.

Where the interaction is measured matters more than its average size, and that
settles a design choice we had made on intuition. It falls sharply with the size
of the set the marginal is measured in: at $|\mathcal{S}|=0$ no sample vanishes
and the median interaction is $0.54$ of the marginal it
perturbs, while at $|\mathcal{S}|=3$ nearly half
vanish and the median is $0.024$, a factor of
23 apart. A
marginal measured against a courier's \emph{current} workload is therefore far
more accurate than one measured against an empty courier. That is
what~\eqref{eq:marginal} does, and it is also why the outer loop recomputes
$C_{rk}$ at every pass: the set of requests a courier holds changes from one
pass to the next, and a marginal measured against requests the courier no longer
holds is measured in the wrong place.

\titledParagraph{Temporal criteria in the grouping cost.}\label{sec:temporal}\label{sec:slack}
In Section~\ref{sec:cost} we argued that a temporal coordinate is the wrong
thing to group by, because it draws together the requests that compete for a
courier rather than those it can serve in turn. That claim can be tested without
reference to our proposed method at all, because two of the six rules differ
in exactly
one respect: $k$-means groups requests by their position, and spatio-temporal
$k$-means adds a scaled time coordinate to the same algorithm. If grouping requests that fall due together
helped, the second would be the better of the two. It is not, and where it
fails is exactly where the argument predicts, as
\tabref{tab:temporal} shows.

\begin{table}[t]\centering
\sbox{\tabbox}{\small\begin{tabular}{@{}l rr r rr@{}}\toprule
& \multicolumn{2}{c}{tardiness} & & \multicolumn{2}{c}{cost}\\
\cmidrule(lr){2-3}\cmidrule(lr){5-6}
instances & $k$-m. & s.-t. & $p$ & $k$-m. & s.-t.\\\midrule
aligned & 575 & 743 & $0.164$ & 6038 & 6251\\
misaligned & 245 & 857 & $0.004$ & 5930 & 6583\\
\bottomrule\end{tabular}}%
\setlength{\tabw}{0.80\textwidth}%
\ifdim\wd\tabbox>\tabw\setlength{\tabw}{\wd\tabbox}\fi
\begin{minipage}{\tabw}
\centering
\caption{$|\mathcal{R}| = 40, 100, 200$ requests, $|\mathcal{K}| = 12, 30, 60$ couriers, $Q_k = 5$, $\varphi = 100$, $\rho = 1$. $k$-means grouping with and without a time coordinate, paired on identical instances. \emph{$k$-m.}\ is $k$-means on request midpoints; \emph{s.-t.}\ is spatio-temporal $k$-means, the same algorithm carrying the extra time coordinate of Section~\ref{sec:setup} at temporal weight $\omega=1$, so the two differ in that coordinate alone. Misaligned instances are those where spatial and temporal structure disagree, so the time coordinate carries information the spatial one does not.}
\label{tab:temporal}
\small\begin{tabular*}{\tabw}{@{\extracolsep{\fill}}l rr r rr@{}}\toprule
& \multicolumn{2}{c}{tardiness} & & \multicolumn{2}{c}{cost}\\
\cmidrule(lr){2-3}\cmidrule(lr){5-6}
instances & $k$-m. & s.-t. & $p$ & $k$-m. & s.-t.\\\midrule
aligned & 575 & 743 & $0.164$ & 6038 & 6251\\
misaligned & 245 & 857 & $0.004$ & 5930 & 6583\\
\bottomrule\end{tabular*}
\end{minipage}
\end{table}

On instances where space and time disagree, adding the time coordinate
multiplies tardiness by $3.5$ ($p=0.004$) and raises total cost by
$11.4\%$ ($p=0.008$). On instances where the two agree
the effect is not significant ($p=0.164$), which is what one would
expect: there the time coordinate merely repeats what the spatial one already
says, so it can do little harm and no good. The published remedy for time
windows in a clustering rule therefore makes the rule worse precisely when it is
supposed to help, and the reason is the one given in
Section~\ref{sec:cost}: a courier serves its group in sequence, so requests
that fall due together contend for it.

The time coordinate carries a weight, so the natural objection is that the
weight was set badly. \figrefp{fig:sensitivity}{c} sweeps the
temporal weight $\omega$ of Section~\ref{sec:setup} over 8 values. No
setting helps by more than noise. The best of them in the rule's published form
is the smallest weight tried, $\omega=0.125$, at
$-1.0\%$ against $\omega=0$ on
7 of 12 instances
($p=0.074$). From $\omega=1$ upward every weight is
significantly worse, and the cost climbs to $+18\%$ at
$\omega=8$ with tardiness climbing alongside it, from
286 to 871.
Carried by our descent the same rule is flat to within 4\%, which is
the descent repairing the grouping rather than the weight becoming useful.

That was the crude way of putting time into a grouping cost. The sophisticated one
concedes what Section~\ref{sec:cost} argues, that minimising a temporal distance
groups the requests that compete for a courier, and introduces time another
way: it
charges each pair of requests the courier cannot chain, by
$g_{\mathrm{sum}}$ or $g_{\min}$ of~\eqref{eq:chaincharge}. That is the
criterion most likely to be proposed in place of the marginal cost, so it is the
one worth measuring. Like the marginal cost of~\eqref{eq:marginal}, this is a
grouping cost, and it is used in its place, so the comparison is between two
answers to the same question. It is built and measured exactly as the six
published rules are, on the same 12 instances.

The chaining slack of Section~\ref{sec:cost} is directional.
$\operatorname{slack}(r\!\to\!r')$ is the spare time left if the courier
serves $r$ and then $r'$, and $\operatorname{slack}(r'\!\to\!r)$ is a
value computed in the other direction. A grouping cost charges a pair rather than an ordered pair, so
the two directions have to be combined into one, and there are two ways to do it.
Summing them charges a pair when \emph{either} order fails; taking their
minimum charges it only when \emph{both} do. The second is the correct reading,
since the courier chooses the best order, and both are measured below.

The two symmetrisations are not variants of one quantity, and correlating each
against the temporal separation $|\tau_r-\tau_{r'}|$ of
Section~\ref{sec:cost} shows why. Taken over all pairs of requests in these
instances, $g_{\mathrm{sum}}$ of~\eqref{eq:chaincharge} correlates $+0.95$ with
separation: it is, to a good approximation, a measure of how far apart in time
two requests are, so minimising it groups requests that fall due together, which
is what Section~\ref{sec:cost} argues against. $g_{\min}$ correlates
$-0.74$, of the opposite sign, scoring requests far apart in
time as the compatible ones. The correction does what it claims.
\tabref{tab:slack} reports what each symmetrisation
costs.

\begin{table}[t]\centering
\sbox{\tabbox}{\small\begin{tabular}{@{}l@{\;} r@{\;\;}r@{\;\;}r@{\;\;} r@{\;\;} r@{}}\toprule
grouping cost & travel & tard. & cour. & cost & vs.\ ours\\\midrule
slack, sum & 2221 & 1387 & 15.2 & 5133 & $+114.8\%$\\
slack, min & 2218 & 1673 & 16.7 & 5558 & $+134.3\%$\\
slack, sum $+$ desc. & 1291 & 11 & 14.7 & 2768 & $+17.2\%$\\
slack, min $+$ desc. & 1245 & 6 & 15.2 & 2776 & $+16.5\%$\\
best published $+$ desc. & 1184 & 13 & 14.7 & 2664 & $+12.6\%$\\
\midrule
marginal cost (ours) & 873 & 42 & 14.3 & 2349 & ---\\
\bottomrule\end{tabular}}%
\setlength{\tabw}{0.80\textwidth}%
\ifdim\wd\tabbox>\tabw\setlength{\tabw}{\wd\tabbox}\fi
\begin{minipage}{\tabw}
\centering
\caption{$|\mathcal{R}| = 40, 100$ requests, $|\mathcal{K}| = 12, 30$ couriers, $Q_k = 5$, $\varphi = 100$, $\rho = 1$. Grouping by pairwise chaining slack, in both symmetrisations, against the marginal cost. \emph{Sum} charges both orders of a pair, \emph{min} only the better one. Each is reported as published and with our descent, and all are routed by the same exact router.}
\label{tab:slack}
\small\begin{tabular*}{\tabw}{@{\extracolsep{\fill}}l@{\;} r@{\;\;}r@{\;\;}r@{\;\;} r@{\;\;} r@{}}\toprule
grouping cost & travel & tard. & cour. & cost & vs.\ ours\\\midrule
slack, sum & 2221 & 1387 & 15.2 & 5133 & $+114.8\%$\\
slack, min & 2218 & 1673 & 16.7 & 5558 & $+134.3\%$\\
slack, sum $+$ desc. & 1291 & 11 & 14.7 & 2768 & $+17.2\%$\\
slack, min $+$ desc. & 1245 & 6 & 15.2 & 2776 & $+16.5\%$\\
best published $+$ desc. & 1184 & 13 & 14.7 & 2664 & $+12.6\%$\\
\midrule
marginal cost (ours) & 873 & 42 & 14.3 & 2349 & ---\\
\bottomrule\end{tabular*}
\end{minipage}
\end{table}

Nonetheless, correcting the symmetrisation does not rescue the criterion. As published the
minimum is if anything the worse of the two
($+134.3\%$ against ours, versus
$+114.8\%$ for the sum), and once
both are given our descent they are statistically indistinguishable from each
other ($-0.5\%$, $p=0.733$) and both remain worse than
the marginal cost by $+17.2\%$ and $+16.5\%$
($p<10^{-3}$). They are also worse than the best published rule carrying
the same descent, so neither reaches the rules the idea was meant to improve on.

The components say where the difference sits, and it is not punctuality. Carrying
our descent, the slack rules are the \emph{more} punctual of the two, with
tardiness 6 and 11 against our 42, but they travel
1245 and 1291 against our 873 and open more couriers. At
$\rho=1$ that trade runs about ten to one against them: the better of the two
saves 36 units of lateness and pays 372 units of
extra distance for it. Section~\ref{sec:rho} shows the same trade against the
published rules, across a twentyfold range in $\rho$.

Getting the sign right is therefore not enough, and that is a separate point from
the correlation. A charge on pairs
asks whether two requests can follow one another; what a courier is actually
charged is the cost of a \emph{path} through all of them, and no sum over pairs
recovers it, because what a pair costs depends on what else the courier is
carrying and where the path has reached. The failure is not in how the two
directions were combined, which the minimum repairs, but in charging pairs at
all. That is the quantity \eqref{eq:marginal} evaluates directly.

In the proposed method time does not enter the grouping cost directly, as it does
in both criteria above. It enters through the marginal cost
of~\eqref{eq:marginal}, which prices a request by the lateness the courier would
actually incur on the route it would then take, at the rate $\rho$ the objective
sets. Our proposed method therefore spends on punctuality exactly what the
objective says it is worth. That is why it is the
less punctual of the two here and still the cheaper, and why the same mechanism
reverses when lateness is priced higher: Section~\ref{sec:rho} follows it across
a twentyfold range in $\rho$. A criterion with no $\rho$ in it cannot make that
trade in either direction.

\subsection{Sensitivity to instance conditions}\label{sec:sensitivity}
\titledParagraph{Spatio-temporal misalignment.}\label{sec:alignaxis}
The comparison so far uses the two ends of the alignment axis, $\alpha=1$ and
$\alpha=0$. Misalignment is measured on each instance rather than set by
$\alpha$, so intermediate values place instances along the axis and the margin
can be read as a function of it. \figrefp{fig:sensitivity}{b} sweeps
$\alpha$ over 5 levels with 8
seeds each.

The margin over the strengthened rules grows monotonically with realised
misalignment: the Spearman rank correlation between the two is
$-0.90$ ($p=0.037$), negative because a wider margin
is a more negative cost difference. Our proposed method is cheaper on every
instance at every level ($p=0.008$ at each). The bars on the right axis of the same panel
show the effect without reference to our proposed method, since they compare
only the
two baselines: the tardiness spatio-temporal $k$-means pays for its time
coordinate, relative to plain $k$-means, falls from $3.8\times$ at the
misaligned end to $1.2\times$ where space and time agree. That
ordering is what Section~\ref{sec:cost} predicts. Where the time axis repeats
what the spatial one already says, adding it costs little; where it carries
information the spatial axis does not, minimising distance along it concentrates
the requests that compete for a courier.

One value from this sweep is used later. The adapted Li \& Lim batches sit at a
realised misalignment of $0.61$, beside our $\alpha=0$ setting at the hard end
of the axis, so the smaller margin Section~\ref{sec:lilim} reports on them is
not explained by their being temporally easy.

\titledParagraph{The penalty rate.}\label{sec:rho}
In Section~\ref{sec:cost} we argued that a courier visits its requests one after
another, so two requests whose windows coincide compete for the same
courier-minute, and that grouping by temporal proximity therefore collects
exactly those competing pairs onto one courier. Grouping on the marginal cost
of~\eqref{eq:marginal} instead of on temporal proximity might therefore be
expected to buy punctuality. However, what it buys is distance: travel falls and
lateness does not. One explanation would be
that $\rho=1$ makes lateness too cheap to be worth buying, in which case pricing
it higher should close the gap, so we swept $\rho$ over a twentyfold range. It enters every part of the
method: the exact router sequences with it, the marginal costs
of~\eqref{eq:marginal} are computed with it, and the descent decides with it.
\tabref{tab:rho} reports the sweep, and
\figrefp{fig:sensitivity}{d} plots it beside the tardiness of both methods.

\begin{table}[t]\centering
\sbox{\tabbox}{\small\begin{tabular}{@{}r@{\;\;} r@{\;\;}r@{\;\;} r@{\;\;}r@{\;\;} r@{\;\;} r@{}}\toprule
& \multicolumn{2}{c}{tardiness} & \multicolumn{2}{c}{travel} & & \\
\cmidrule(lr){2-3}\cmidrule(lr){4-5}
$\rho$ & ours & base & ours & base & cost & $p$\\\midrule
$0.5$ & 65.66 & 29.41 & 564 & 696 & $-11.1\%$ & $0.008$\\
$1$ & 14.34 & 7.45 & 606 & 694 & $-9.1\%$ & $0.008$\\
$2$ & 4.46 & 2.82 & 645 & 739 & $-8.7\%$ & $0.008$\\
$5$ & 0.90 & 0.18 & 637 & 740 & $-9.3\%$ & $0.008$\\
$10$ & 0.73 & 0.03 & 644 & 720 & $-8.2\%$ & $0.008$\\
\bottomrule\end{tabular}}%
\setlength{\tabw}{0.80\textwidth}%
\ifdim\wd\tabbox>\tabw\setlength{\tabw}{\wd\tabbox}\fi
\begin{minipage}{\tabw}
\centering
\caption{$|\mathcal{R}| = 40$ requests, $|\mathcal{K}| = 12$ couriers, $Q_k = 5$, $\varphi = 100$, $\rho = 1$. Sweeping the penalty rate $\rho$, which prices earliness and tardiness alike in~\eqref{eq:globalobj}. \emph{Base} is the best of the six rules carrying our descent on each instance. Tardiness is a mean over the instances, so a small value is an average of instances that are exactly punctual and one or two that are not, rather than a small lateness everywhere. The proposed method responds to $\rho$ as it should, and its advantage stays travel at every $\rho$.}
\label{tab:rho}
\small\begin{tabular*}{\tabw}{@{\extracolsep{\fill}}r@{\;\;} r@{\;\;}r@{\;\;} r@{\;\;}r@{\;\;} r@{\;\;} r@{}}\toprule
& \multicolumn{2}{c}{tardiness} & \multicolumn{2}{c}{travel} & & \\
\cmidrule(lr){2-3}\cmidrule(lr){4-5}
$\rho$ & ours & base & ours & base & cost & $p$\\\midrule
$0.5$ & 65.66 & 29.41 & 564 & 696 & $-11.1\%$ & $0.008$\\
$1$ & 14.34 & 7.45 & 606 & 694 & $-9.1\%$ & $0.008$\\
$2$ & 4.46 & 2.82 & 645 & 739 & $-8.7\%$ & $0.008$\\
$5$ & 0.90 & 0.18 & 637 & 740 & $-9.3\%$ & $0.008$\\
$10$ & 0.73 & 0.03 & 644 & 720 & $-8.2\%$ & $0.008$\\
\bottomrule\end{tabular*}
\end{minipage}
\end{table}

As $\rho$ rises from 0.5 to 10, the mean tardiness of our proposed method
falls from 65.7 to 0.7 and its mean travel rises from 564 to
644.
The comparison with the baselines does not move with it. At every $\rho$ our
solutions are the later of the two, by roughly a factor of two while lateness is
still material and by less than a unit once $\rho$ has driven both near zero, and
at every $\rho$ our travel is lower, by $76$ to $132$ units.
The cost advantage is flat across the range, $-11.1\%$ at
$\rho=0.5$ and $-8.2\%$ at $\rho=10$, on every instance at
every $\rho$. That total is the sum of two terms moving in opposite directions
and by very different amounts: measured in the units of
the objective~\eqref{eq:globalobj}, the travel our proposed method saves
outweighs the lateness it gives up for that saving by a factor of 7 to
29 across the sweep. The cost is therefore lower than every one of the
six strengthened rules at every penalty rate, and the gain is a large saving
in travel bought with a small concession in tardiness.

Chainability and compactness are the same property. A group whose requests chain
is one whose courier can leave each request and reach the next without a long
empty run, which is a group with short travel between its stops, and the marginal
cost of~\eqref{eq:marginal} selects for it directly. Punctuality is available to
either method once $\rho$ makes it worth buying, both driving lateness towards
zero by $\rho=10$; what differs is the distance each pays to get there. The
contribution is therefore the grouping criterion and its effect on cost, and not
an improvement in punctuality.

\subsection{The dispatch charge and the fleet}\label{sec:fleetcharge}
\titledParagraph{Behaviour without a dispatch charge.}\label{sec:phizero}
Everything above prices a dispatched courier at $\varphi = 100$. Setting
$\varphi = 0$ removes the fleet decision from the objective: any number of
couriers may be opened at no charge, and the problem reduces to minimising
travel plus lateness. Since the dispatch charge is what no published clustering
rule can respond to, this is the setting in which our advantage over them should
be most at risk, and \tabref{tab:phizero} repeats the whole comparison there.

\begin{table}[t]\centering
\sbox{\tabbox}{\small\begin{tabular}{@{}l rrrr rrrr@{}}
\toprule
& \multicolumn{4}{c}{as published} & \multicolumn{4}{c}{+ descent}\\
\cmidrule(lr){2-5}\cmidrule(lr){6-9}
method & travel & tard. & cour. & cost & travel & tard. & cour. & cost\\
\midrule
\textsc{sweep} & 2163 & 833 & 22.7 & 2996 & 1786 & 8 & 30.0 & 1794\\
\textsc{fisher} & 2487 & 578 & 34.0 & 3064 & 1605 & 5 & 32.4 & 1610\\
$k$-means & 2174 & 410 & 34.0 & 2584 & 1829 & 11 & 31.8 & 1839\\
spatio-temporal $k$-means & 2217 & 800 & 34.0 & 3017 & 1936 & 28 & 32.8 & 1964\\
capacitated $k$-means & 2284 & 449 & 34.0 & 2733 & 1868 & 10 & 32.1 & 1878\\
Ward agglomerative & 2176 & 326 & 34.0 & 2502 & 1867 & 40 & 32.1 & 1907\\
\midrule
\textbf{this paper} & \multicolumn{4}{c}{} & 1095 & 10 & 31.5 & 1106\\
ALNS & \multicolumn{4}{c}{} & 1104 & 40 & 30.8 & 1144\\
\bottomrule\end{tabular}}%
\setlength{\tabw}{0.80\textwidth}%
\ifdim\wd\tabbox>\tabw\setlength{\tabw}{\wd\tabbox}\fi
\begin{minipage}{\tabw}
\centering
\caption{$|\mathcal{R}| = 40, 100, 200$ requests, $|\mathcal{K}| = 12, 30, 60$ couriers, $Q_k = 5$, $\varphi = 0$, $\rho = 1$. The same instances with no dispatch charge, $\varphi = 0$, so cost is travel${}+{}$tardiness alone and the courier count is free. Compare with \tabref{tab:main}.}
\label{tab:phizero}
\small\begin{tabular*}{\tabw}{@{\extracolsep{\fill}}l rrrr rrrr@{}}
\toprule
& \multicolumn{4}{c}{as published} & \multicolumn{4}{c}{+ descent}\\
\cmidrule(lr){2-5}\cmidrule(lr){6-9}
method & travel & tard. & cour. & cost & travel & tard. & cour. & cost\\
\midrule
\textsc{sweep} & 2163 & 833 & 22.7 & 2996 & 1786 & 8 & 30.0 & 1794\\
\textsc{fisher} & 2487 & 578 & 34.0 & 3064 & 1605 & 5 & 32.4 & 1610\\
$k$-means & 2174 & 410 & 34.0 & 2584 & 1829 & 11 & 31.8 & 1839\\
spatio-temporal $k$-means & 2217 & 800 & 34.0 & 3017 & 1936 & 28 & 32.8 & 1964\\
capacitated $k$-means & 2284 & 449 & 34.0 & 2733 & 1868 & 10 & 32.1 & 1878\\
Ward agglomerative & 2176 & 326 & 34.0 & 2502 & 1867 & 40 & 32.1 & 1907\\
\midrule
\textbf{this paper} & \multicolumn{4}{c}{} & 1095 & 10 & 31.5 & 1106\\
ALNS & \multicolumn{4}{c}{} & 1104 & 40 & 30.8 & 1144\\
\bottomrule\end{tabular*}
\end{minipage}
\end{table}

The advantage over the clustering rules does not require the dispatch charge,
and is larger without one. Against the six rules as published our proposed method is
$-49.9\%$ (18/18 instances, $p<10^{-3}$), and
against the same rules carrying our descent $-26.5\%$
(18/18, $p<10^{-3}$), against $-12.1\%$ for the
same comparison at $\varphi = 100$. The apparent strengthening is largely
arithmetic: at $\varphi = 100$ every method pays a fixed term of $\varphi$ times
the couriers it opens, 23 on average over these instances, a large
common addend that compresses every relative difference. In absolute terms nothing changes, and the travel column shows it: our proposed
method is close to ALNS there and the rules are not.

The courier counts separate the methods by whether the fleet size is a decision
at all. The number of groups is an input to every published rule, so each of the
6 dispatches the same couriers whichever charge applies: the counts in the \emph{as published} block of \tabref{tab:main}
and of \tabref{tab:phizero} are identical. Our proposed method opens
8.3 more couriers when the charge is lifted and ALNS
7.8 more, both responding to what the objective now prices. The
strengthened rules respond as well, by 3.7 to 7.3
couriers, because the descent they carry minimises the objective, and the
objective charges $\varphi$ for every courier opened.

This is the form in which the fleet result holds. It is a claim about
responsiveness to $\varphi$ and it separates our proposed method from the
published rules, not from ALNS. It is not a claim about distance from the
capacity floor of Section~\ref{sec:problemFormulation}: \textsc{sweep} fills
each courier to $Q_k$ in angular order, so it opens
$\lceil|\mathcal{R}|/\max_k Q_k\rceil$ couriers exactly, sitting on the floor
at $\varphi=100$ and at $\varphi=0$ alike.

Against ALNS the two are not statistically distinguishable: pooled
$-0.5\%$ with 11 wins in 18, $p=0.167$. That
pooled figure hides a trend with size, $+9.6\%$ at $|\mathcal{R}|=40$, $-5.8\%$ at $|\mathcal{R}|=100$, $-5.2\%$ at $|\mathcal{R}|=200$, so ALNS is ahead on the
smallest instances and behind on the larger ones.
Under a dispatch charge the decomposition beats ALNS at matched
wall-clock and ALNS does not catch up given far longer
(Section~\ref{sec:crossover}); with couriers free it is a tie at matched
wall-clock, and a search given more time would be expected to pass us, since
what remains in the objective is exactly the route quality a search over
complete solutions is good at.

Against every published clustering rule, in its own form and strengthened with
our improvement step, our proposed method is cheaper on all 18 instances
at $\varphi = 100$ and on all 18 at $\varphi = 0$, and against ALNS it is
ahead under a dispatch charge and level without one. Both comparisons use the
three problem sizes of \tabref{tab:main}. Section~\ref{sec:scaling} takes the
one against the clustering rules to $1000$ requests and
$300$ couriers, where most of our runtime is work that decomposes by
courier: our proposed method returns in
50\,s against
821\,s for the strongest clustering baseline, at a cost
34\% below
it.

\titledParagraph{The charge swept.}\label{sec:phisweep}
Section~\ref{sec:rules} prices a dispatched courier at $\varphi=100$ and
Section~\ref{sec:phizero} at $\varphi=0$. Sweeping $\varphi$ over
6 levels from $0$ to $400$ at
$|\mathcal{R}|=40$ gives the shape between them;
\figrefp{fig:sensitivity}{a} plots the result. The margin over the
strengthened rules is monotone and \emph{largest} where the charge is
smallest, $-15.1\%$ at $\varphi=0$ against
$-4.2\%$ at $\varphi=400$, which is the compression
effect of Section~\ref{sec:phizero} seen across the whole range rather than at
its ends: a term every method pays alike grows with $\varphi$ and shrinks every
ratio. Our fleet falls from 11.2
couriers at $\varphi=0$ to the capacity floor of $8$ by
$\varphi=100$ and stays there, while no rule in its published form moves
at any charge. Against ALNS the sign changes
inside the range: we are behind at $\varphi=0$ and ahead from
$\varphi=50$ upwards, which locates the crossover the earlier
sections describe.

\begin{table}[t]\centering
\sbox{\tabbox}{\small\begin{tabular}{@{}rr rrr r@{}}\toprule
& & \multicolumn{3}{c}{vs.\ rules $+$ descent} & vs.\ ALNS\\
\cmidrule(lr){3-5}\cmidrule(lr){6-6}
$|\mathcal{K}|$ & $|\mathcal{R}|$ & mean & wins & $p$ & mean\\\midrule
$6$ & $20$ & $-4.1\%$ & 6/8 & $0.055$ & $+1.7\%$\\
$9$ & $32$ & $-6.5\%$ & 7/8 & $0.039$ & $-1.1\%$\\
$12$ & $40$ & $-9.1\%$ & 8/8 & $0.008$ & $-4.3\%$\\
$18$ & $60$ & $-12.2\%$ & 8/8 & $0.008$ & $-10.7\%$\\
\bottomrule\end{tabular}}%
\setlength{\tabw}{0.80\textwidth}%
\ifdim\wd\tabbox>\tabw\setlength{\tabw}{\wd\tabbox}\fi
\begin{minipage}{\tabw}
\centering
\caption{$|\mathcal{R}| = 20, 32, 40, 60$ requests, $|\mathcal{K}| = 6, 9, 12, 18$ couriers, $Q_k = 5$, $\varphi = 100$, $\rho = 1$. The advantage widens with the fleet. The design rule of Section~\ref{sec:setup} is held fixed, so the workload grows with the fleet. Paired against the best of the six rules carrying our descent, and against ALNS at matched wall-clock, at $\varphi=100$. Negative favours the proposed method.}
\label{tab:fleet}
\small\begin{tabular*}{\tabw}{@{\extracolsep{\fill}}rr rrr r@{}}\toprule
& & \multicolumn{3}{c}{vs.\ rules $+$ descent} & vs.\ ALNS\\
\cmidrule(lr){3-5}\cmidrule(lr){6-6}
$|\mathcal{K}|$ & $|\mathcal{R}|$ & mean & wins & $p$ & mean\\\midrule
$6$ & $20$ & $-4.1\%$ & 6/8 & $0.055$ & $+1.7\%$\\
$9$ & $32$ & $-6.5\%$ & 7/8 & $0.039$ & $-1.1\%$\\
$12$ & $40$ & $-9.1\%$ & 8/8 & $0.008$ & $-4.3\%$\\
$18$ & $60$ & $-12.2\%$ & 8/8 & $0.008$ & $-10.7\%$\\
\bottomrule\end{tabular*}
\end{minipage}
\end{table}
\titledParagraph{The fleet size.}\label{sec:fleetsweep}
The design rule of Section~\ref{sec:setup} ties the request count to the
fleet, so enlarging $|\mathcal{K}|$ enlarges the instance with it and the
workload per courier is held fixed. \tabref{tab:fleet} sweeps
$|\mathcal{K}|$ from 6 to 18.

Both margins move monotonically with the fleet: against the strengthened rules
from $-4.1\%$ at $|\mathcal{K}|=6$ to
$-12.2\%$ at $|\mathcal{K}|=18$, and against
ALNS from $+1.7\%$ to
$-10.7\%$, crossing from behind to ahead. The reason is the
fleet decision itself: the number of couriers a solution may open grows with
$|\mathcal{K}|$, so there is more for the assignment program to decide and
more for a method that cannot decide it to lose.

\subsection{Scaling, and where the work sits}\label{sec:scaling}
\begin{table}[t]\centering
\sbox{\tabbox}{\small\begin{tabular}{@{}r l rrr r rr@{}}\toprule
$|\mathcal{R}|$ & method & travel & tard. & cour. & cost & time & assign.\\\midrule
$100$ & ours & 1117 & 4 & 20.5 & 3171 & 7\,s & 6\,s\\
 & baseline & 1936 & 20 & 26.5 & 4606 & 1\,s & ---\\
 & ALNS & 1589 & 25 & 20.5 & 3664 & 11\,s & ---\\
\addlinespace[2pt]
$200$ & ours & 1847 & 1 & 40.5 & 5898 & 18\,s & 13\,s\\
 & baseline & 3246 & 4 & 47.5 & 8000 & 8\,s & ---\\
 & ALNS & 3235 & 59 & 40.5 & 7344 & 19\,s & ---\\
\addlinespace[2pt]
$400$ & ours & 3527 & 14 & 81.0 & 11642 & 31\,s & 20\,s\\
 & baseline & 7115 & 41 & 104.0 & 17556 & 36\,s & ---\\
 & ALNS & 6753 & 158 & 80.5 & 14961 & 90\,s & ---\\
\addlinespace[2pt]
$1000$ & ours & 7878 & 24 & 201.0 & 28002 & 50\,s & 20\,s\\
 & baseline & 16448 & 63 & 262.0 & 42711 & 821\,s & ---\\
 & ALNS & 23518 & 2602 & 200.5 & 46170 & 1483\,s & ---\\
\bottomrule\end{tabular}}%
\setlength{\tabw}{0.80\textwidth}%
\ifdim\wd\tabbox>\tabw\setlength{\tabw}{\wd\tabbox}\fi
\begin{minipage}{\tabw}
\centering
\caption{$|\mathcal{R}| = 100, 200, 400, 1000$ requests, $|\mathcal{K}| = 30, 60, 120, 300$ couriers, $Q_k = 5$, $\varphi = 100$, $\rho = 1$. Two misaligned instances at each size; those at $|\mathcal{R}| = 100$ and $200$ are also in \tabref{tab:main}, those at $400$ and $1000$ are not. Scaling of the proposed method against the strongest clustering baseline (capacitated $k$-means with our descent) on the same instances. \emph{Assign.} is the part of our time that does not divide across cores.}
\label{tab:scale}
\small\begin{tabular*}{\tabw}{@{\extracolsep{\fill}}r l rrr r rr@{}}\toprule
$|\mathcal{R}|$ & method & travel & tard. & cour. & cost & time & assign.\\\midrule
$100$ & ours & 1117 & 4 & 20.5 & 3171 & 7\,s & 6\,s\\
 & baseline & 1936 & 20 & 26.5 & 4606 & 1\,s & ---\\
 & ALNS & 1589 & 25 & 20.5 & 3664 & 11\,s & ---\\
\addlinespace[2pt]
$200$ & ours & 1847 & 1 & 40.5 & 5898 & 18\,s & 13\,s\\
 & baseline & 3246 & 4 & 47.5 & 8000 & 8\,s & ---\\
 & ALNS & 3235 & 59 & 40.5 & 7344 & 19\,s & ---\\
\addlinespace[2pt]
$400$ & ours & 3527 & 14 & 81.0 & 11642 & 31\,s & 20\,s\\
 & baseline & 7115 & 41 & 104.0 & 17556 & 36\,s & ---\\
 & ALNS & 6753 & 158 & 80.5 & 14961 & 90\,s & ---\\
\addlinespace[2pt]
$1000$ & ours & 7878 & 24 & 201.0 & 28002 & 50\,s & 20\,s\\
 & baseline & 16448 & 63 & 262.0 & 42711 & 821\,s & ---\\
 & ALNS & 23518 & 2602 & 200.5 & 46170 & 1483\,s & ---\\
\bottomrule\end{tabular*}
\end{minipage}
\end{table}

\tabref{tab:scale} runs the comparison at four sizes up
to $|\mathcal{R}| = 1000$ requests and
$|\mathcal{K}| = 300$ couriers,
five times the largest instance in
\tabref{tab:main}. At that size our proposed method returns a
solution in 50\,s against 821\,s for the strongest
clustering baseline, and it is ahead on every component rather than on the
total alone: travel 7878 against 16448,
lateness 24 against 63, and
201 couriers against 262, for a cost
34\% below the
baseline's. Of that time, 30\,s is spent in work that is independent per
courier, namely marginal costs, descent screening and exact routing, and
20\,s in the assignment program, which is not. Were the first term to
divide perfectly and the second not at all, the whole would still take 20\,s on
any number of cores: by Amdahl's law the attainable speedup is the total divided
by the serial part, $50/20\approx2.5\times$. That figure is a ceiling computed
from a single-core measurement, not a speedup we observed: every runtime
reported in this paper is single-core and no experiment exploited the parallel
structure. A real implementation would fall short of it, since the
synchronisation points of
Section~\ref{sec:parallel} and the cost of distributing the marginal costs each
take back part of it. What the decomposition delivers is a workload whose
expensive part is per-courier; converting that into wall-clock on a parallel
machine is left to future work. The
candidate list of Section~\ref{sec:assign} contributes as well, by making
every step linear in $|\mathcal{R}|$.

The ALNS rows of \tabref{tab:scale} are not a comparison at equal wall-clock,
as its time column shows. ALNS is given our runtime on the same instance as its
budget, but a budget
can only be tested between iterations, so the search runs to the end of whichever
ruin-and-recreate pass is in flight when the budget expires. At
$100$ and $200$ requests a pass is short enough that this
costs almost nothing and ALNS keeps to the budget. At $400$ it takes
three times what it was allowed, and at
$1000$, where one pass over the whole plan is itself longer
than the budget, it takes 1483\,s against the
50\,s it was allowed, a factor of
30. Even
with that much extra time its solution is far worse on the two components a
search should be good at, travel 23518 against
7878 and lateness 2602 against
24, while it opens the same fleet as we do,
200 couriers against 201. A search over complete solutions
must repair a thousand-request plan at every step, while the decomposition never
sequences more than $Q_k$ requests at once.

\subsection{Benchmark instances and the boundary of the advantage}\label{sec:external}
\titledParagraph{Adapted benchmark instances.}\label{sec:lilim}
The instances above come from our own generator, so we repeat the comparison
analysis on batches drawn from the Li \& Lim benchmark~\citep{SINTEFLiLimBenchmark}, adapted to
this setting as Section~\ref{sec:setup} describes. These have window widths,
spatial layouts and release patterns we did not choose, and they sit at the
misaligned end of the alignment axis throughout. \tabref{tab:lilim} reports
them by family.

\begin{table}[t]\centering
\sbox{\tabbox}{\small\begin{tabular}{@{}l rrrr@{}}\toprule
& ours & best of six & $+$ descent & ALNS\\\midrule
\multicolumn{5}{@{}l}{\emph{cost, by family}}\\
\quad\texttt{lc201} & 1632 & 1799 & 1663 & 1676\\
\quad\texttt{lc202} & 1581 & 1756 & 1631 & 1556\\
\quad\texttt{lr201} & 1779 & 1875 & 1814 & 1833\\
\quad\texttt{lr202} & 1977 & 2208 & 2070 & 2060\\
\quad\texttt{lrc101} & 2449 & 2827 & 2463 & 2475\\
\quad\texttt{lrc102} & 2271 & 2787 & 2258 & 2247\\
\quad\texttt{lrc201} & 2151 & 2517 & 2227 & 2161\\
\quad\texttt{lrc202} & 2125 & 2546 & 2231 & 2101\\
\addlinespace[2pt]\midrule
\multicolumn{5}{@{}l}{\emph{pooled, by component}}\\
\quad travel & 1066 & 1249 & 1084 & 1076\\
\quad tardiness & 47 & 140 & 44 & 54\\
\quad couriers & 8.8 & 9.0 & 9.2 & 8.8\\
\quad \textbf{cost} & \textbf{1996} & 2289 & 2045 & 2014\\
\bottomrule\end{tabular}}%
\setlength{\tabw}{0.80\textwidth}%
\ifdim\wd\tabbox>\tabw\setlength{\tabw}{\wd\tabbox}\fi
\begin{minipage}{\tabw}
\centering
\caption{$|\mathcal{R}| = 40$ requests, $|\mathcal{K}| = 12$ couriers, $Q_k = 5$, $\varphi = 100$, $\rho = 1$. Adapted Li \& Lim batches. \emph{Upper block}: mean cost by family. \emph{Lower block}: the same batches pooled and split into the terms of~\eqref{eq:globalobj}. \emph{Best of six} takes the cheapest published rule on each instance, \emph{+ descent} the cheapest after our improvement step.}
\label{tab:lilim}
\small\begin{tabular*}{\tabw}{@{\extracolsep{\fill}}l rrrr@{}}\toprule
& ours & best of six & $+$ descent & ALNS\\\midrule
\multicolumn{5}{@{}l}{\emph{cost, by family}}\\
\quad\texttt{lc201} & 1632 & 1799 & 1663 & 1676\\
\quad\texttt{lc202} & 1581 & 1756 & 1631 & 1556\\
\quad\texttt{lr201} & 1779 & 1875 & 1814 & 1833\\
\quad\texttt{lr202} & 1977 & 2208 & 2070 & 2060\\
\quad\texttt{lrc101} & 2449 & 2827 & 2463 & 2475\\
\quad\texttt{lrc102} & 2271 & 2787 & 2258 & 2247\\
\quad\texttt{lrc201} & 2151 & 2517 & 2227 & 2161\\
\quad\texttt{lrc202} & 2125 & 2546 & 2231 & 2101\\
\addlinespace[2pt]\midrule
\multicolumn{5}{@{}l}{\emph{pooled, by component}}\\
\quad travel & 1066 & 1249 & 1084 & 1076\\
\quad tardiness & 47 & 140 & 44 & 54\\
\quad couriers & 8.8 & 9.0 & 9.2 & 8.8\\
\quad \textbf{cost} & \textbf{1996} & 2289 & 2045 & 2014\\
\bottomrule\end{tabular*}
\end{minipage}
\end{table}

Pooled over the 24 benchmark batches our proposed method is
$-12.2\%$ against the best published rule (24 wins,
$p<10^{-3}$), a figure that coincides to three significant figures with an
unrelated one, the $-12.2\%$ against \emph{strengthened}
rules at $|\mathcal{K}|=18$ in \tabref{tab:fleet}, which is
measured on generated instances; the two are different comparisons on different
data and agree by chance. Against the best of those rules carrying our descent
the margin here is $-2.4\%$
(18 wins, $p=0.001$). Against ALNS at matched
wall-clock it is $-0.9\%$ (15/24,
$p=0.101$), a tie at this number of instances.
The lower block of \tabref{tab:lilim} gives the components, which locate
that margin: travel and lateness are nearly level with the best strengthened rule
and with ALNS, and most of the difference comes from the fleet size.

Both figures are measured at $|\mathcal{R}| = 40$:
$-8.9\%$ on our generated instances against
$-2.4\%$ on the benchmark batches. Instance size is therefore not
what makes the margin smaller here.
Section~\ref{sec:instancespace} identifies it.

\titledParagraph{Instance structure and the size of the advantage.}\label{sec:instancespace}
The margin on the benchmark batches is smaller than on the instances generated
by the design rule of Section~\ref{sec:setup}. Whether that rule yields
conditions favourable to the proposed method is a question about instance
structure, and we answer it by moving the generator along the axes on which the
two datasets differ and watching the margin. Three axes were tried.
\tabref{tab:instancespace} carries the two swept here; the
alignment axis was swept already in Section~\ref{sec:alignaxis}.
Two do not explain the gap and one
does. The two failures are reported because they say what the effect is
\emph{not}.

\begin{table}[t]\centering
\sbox{\tabbox}{\small\begin{tabular}{@{}l r@{\;\;} r@{\;\;} r@{}}\toprule
setting & property & $+$ desc. & vs.\ ALNS\\\midrule
\multicolumn{4}{@{}l}{\emph{window width: courier-time ratio}}\\
\quad width $45$ & $1.77$ & $-9.1\%$ & $-4.1\%$\\
\quad width $90$ & $2.05$ & $-13.4\%$ & $-5.9\%$\\
\quad width $180$ & $2.61$ & $-14.9\%$ & $-4.2\%$\\
\quad width $360$ & $3.75$ & $-14.3\%$ & $-3.6\%$\\
\quad width $720$ & $6.01$ & $-14.9\%$ & $-4.1\%$\\
\addlinespace[2pt]
\multicolumn{4}{@{}l}{\emph{delivery placement: locality}}\\
\quad local & $0.85$ & $-9.1\%$ & $-3.6\%$\\
\quad cross & $0.03$ & $-2.0\%$ & $-0.2\%$\\
\midrule
benchmark (Li \& Lim) & $0.29$ & $-2.4\%$ & $-0.9\%$\\
\bottomrule\end{tabular}}%
\setlength{\tabw}{0.80\textwidth}%
\ifdim\wd\tabbox>\tabw\setlength{\tabw}{\wd\tabbox}\fi
\begin{minipage}{\tabw}
\centering
\caption{$|\mathcal{R}| = 40$ requests, $|\mathcal{K}| = 12$ couriers, $Q_k = 5$, $\varphi = 100$, $\rho = 1$. Moving our generator towards the benchmark, one axis at a time. Each block reports the measured instance property and the margin against the best of the six rules carrying our descent, and against ALNS. The benchmark row is the adapted Li \& Lim batches at the same size.}
\label{tab:instancespace}
\small\begin{tabular*}{\tabw}{@{\extracolsep{\fill}}l r@{\;\;} r@{\;\;} r@{}}\toprule
setting & property & $+$ desc. & vs.\ ALNS\\\midrule
\multicolumn{4}{@{}l}{\emph{window width: courier-time ratio}}\\
\quad width $45$ & $1.77$ & $-9.1\%$ & $-4.1\%$\\
\quad width $90$ & $2.05$ & $-13.4\%$ & $-5.9\%$\\
\quad width $180$ & $2.61$ & $-14.9\%$ & $-4.2\%$\\
\quad width $360$ & $3.75$ & $-14.3\%$ & $-3.6\%$\\
\quad width $720$ & $6.01$ & $-14.9\%$ & $-4.1\%$\\
\addlinespace[2pt]
\multicolumn{4}{@{}l}{\emph{delivery placement: locality}}\\
\quad local & $0.85$ & $-9.1\%$ & $-3.6\%$\\
\quad cross & $0.03$ & $-2.0\%$ & $-0.2\%$\\
\midrule
benchmark (Li \& Lim) & $0.29$ & $-2.4\%$ & $-0.9\%$\\
\bottomrule\end{tabular*}
\end{minipage}
\end{table}

\titledParagraph{Not the alignment.} Section~\ref{sec:alignaxis} places the
benchmark at a realised misalignment of $0.61$, beside our hardest setting,
where our margin is at its largest. If anything the benchmark sits where the
method should do best.

\titledParagraph{Not the time pressure.} Widening the time windows moves the
courier-time ratio from $1.77$ to
$6.01$, across and beyond the benchmark's $2.61$. The
margin does not shrink; it \emph{grows}, from $-9.1\%$ to
$-14.9\%$. At the width whose courier-time ratio matches the
benchmark exactly the margin is $-14.9\%$,
six times the benchmark's. How much time the fleet has is not the explanation,
and we had expected it to be.

\titledParagraph{The spatial structure.} What does explain it is where a
request's delivery lies. Our default generator puts the delivery near its own
pickup, so the locality defined in Section~\ref{sec:setup}, the fraction of
requests whose pickup and delivery lie within $10$ of each other, is
$0.85$. Sending the delivery to a different hub gives
$0.03$, and the benchmark sits between the two at $0.29$. The margin
against the strengthened rules collapses from $-9.1\%$ to
$-2.0\%$ (7/8, $p=0.109$), and against
ALNS from $-3.6\%$ to $-0.2\%$, within noise of
the $-2.4\%$ and $-0.9\%$ measured on the benchmark
itself.

The mechanism is the one Section~\ref{sec:cost} argues for, seen at its
boundary. When a
delivery sits near its pickup, a request is effectively a point, the freedom
lies in which points travel together, and what a group costs is exactly the
chaining the marginal cost measures and a centroid cannot. When the delivery is
across the city, every request is a long mandatory traversal, the route is
dominated by edges no grouping can avoid, and the choice of partition matters
much less: any sensible grouping followed by a good descent reaches a similar
place. The benchmark lies towards that end, which is why our advantage there is
real but small.

Courier dispatch sits at the other end of the same axis. In meal delivery, the
setting whose operational conditions this paper
takes as its model~\citep{reyes2018meal, yildiz2019provably,
ulmer2021restaurant}, and in comparable work such as pharmacy delivery, a
customer orders from a vendor already close by and the platform serves the order
within a bounded radius, so a request's pickup and its delivery are near each
other. That is why the generator of Section~\ref{sec:setup} places each
delivery near its own pickup, and it is the regime in which the grouping
criterion rather than the unavoidable edges decides what a solution costs. The
boundary is stated here because the margin is not uniform across instance
structures, and because the structures in which it is largest are the ones the
application presents.

\subsection{Comparison with the metaheuristic reference}

\tabref{tab:main} also lists an adaptive large neighbourhood search under the
budget of Section~\ref{sec:setup}, our own wall-clock on each instance with a
ten-second floor. Under that budget
the decomposition is $-14.2\%$ against it, winning 18 of
18 instances ($p<10^{-3}$), and \tabref{tab:bysize} splits that by
size. The difference is not the fleet, since the two dispatch within
3.7\% of each other at every size. It is travel, where our margin
widens as the instances grow: $4\%$ at $|\mathcal{R}|=40$, $37\%$ at $|\mathcal{R}|=100$, $47\%$ at $|\mathcal{R}|=200$.
The explanation is the budget rather than the algorithm. A ruin-and-recreate
step reinserts requests into a plan spanning the whole instance, so the work per
improvement grows with $|\mathcal{R}|$, while the decomposition never sequences
more than $Q_k$ requests at a time and solves each of those exactly. Given our
wall-clock, ALNS has fewer useful moves the larger the instance is.
Section~\ref{sec:crossover} removes the budget restriction and measures how
much more time it needs.

\titledParagraph{Time for the reference to match.}\label{sec:crossover}
\begin{table}[t]\centering
\sbox{\tabbox}{\small\begin{tabular}{@{}r@{\;\;} l@{\;\;} r@{\;\;}r@{\;\;}r@{\;\;} r@{\;\;} r@{}}\toprule
$|\mathcal{R}|$ & & travel & tard. & cour. & cost & time\\\midrule
$100$ & ours & 1117 & 4 & 20.5 & 3171 & 7.6\,s\\
 & ALNS & 1252 & 6 & 20.5 & 3307 ($+4\%$) & 300\,s\\
\addlinespace[2pt]
$200$ & ours & 1847 & 1 & 40.5 & 5898 & 18.1\,s\\
 & ALNS & 2454 & 25 & 41.5 & 6629 ($+12\%$) & 300\,s\\
\addlinespace[2pt]
$400$ & ours & 3527 & 14 & 81.0 & 11642 & 37.7\,s\\
 & ALNS & 6129 & 117 & 81.0 & 14345 ($+23\%$) & 300\,s\\
\bottomrule\end{tabular}}%
\setlength{\tabw}{0.80\textwidth}%
\ifdim\wd\tabbox>\tabw\setlength{\tabw}{\wd\tabbox}\fi
\begin{minipage}{\tabw}
\centering
\caption{$|\mathcal{R}| = 100, 200, 400$ requests, $|\mathcal{K}| = 30, 60, 120$ couriers, $Q_k = 5$, $\varphi = 100$, $\rho = 1$. ALNS is stopped as soon as its incumbent reaches the cost of the proposed method, or after 300\,s. It never reached it. Components are shown for both, so the residual gap can be attributed. Two misaligned instances per size, each searched by ALNS with two seeds; the ALNS row is the worse of the two seeds on each instance, the only one whose components were recorded. The better seed is $+4\%$, $+9\%$ and $+21\%$ above the proposed method at the three sizes.}
\label{tab:crossover}
\small\begin{tabular*}{\tabw}{@{\extracolsep{\fill}}r@{\;\;} l@{\;\;} r@{\;\;}r@{\;\;}r@{\;\;} r@{\;\;} r@{}}\toprule
$|\mathcal{R}|$ & & travel & tard. & cour. & cost & time\\\midrule
$100$ & ours & 1117 & 4 & 20.5 & 3171 & 7.6\,s\\
 & ALNS & 1252 & 6 & 20.5 & 3307 ($+4\%$) & 300\,s\\
\addlinespace[2pt]
$200$ & ours & 1847 & 1 & 40.5 & 5898 & 18.1\,s\\
 & ALNS & 2454 & 25 & 41.5 & 6629 ($+12\%$) & 300\,s\\
\addlinespace[2pt]
$400$ & ours & 3527 & 14 & 81.0 & 11642 & 37.7\,s\\
 & ALNS & 6129 & 117 & 81.0 & 14345 ($+23\%$) & 300\,s\\
\bottomrule\end{tabular*}
\end{minipage}
\end{table}

Given the same wall-clock the proposed method is ahead, and the natural objection
is that ALNS would win with more time. \tabref{tab:crossover} tests
that directly, by stopping the search the moment its incumbent reaches our
cost. Over every instance and both seeds it never did, inside a budget of
300\,s against our 7--38\,s, and the shortfall
widens with size, reaching $+23\%$ at
$|\mathcal{R}| = 400$, and $+21\%$ for the better of the two seeds.

The components say what the residual is, and the obvious guess about it is
wrong. One might expect a ruin-and-recreate search to be held
back by the fleet. Closing a courier means relocating every one of its requests,
and every intermediate step raises the objective, because a route with one
request left still pays $\varphi$ in full. No standard removal rule performs that
transformation in a single move. That is not what happens. At $|\mathcal{R}| = 400$ ALNS dispatches
81.0 couriers against our
81.0, so it sizes the fleet as we do. The gap is
travel, 2601 units of it, about
96\% of the shortfall.

The reason is the size of the neighbourhood rather than its shape, and it is the
one given above: each ruin-and-recreate step works on a plan spanning the whole
instance, so three hundred seconds buys proportionally fewer useful moves the
larger that instance is, which is why the shortfall widens rather than closes. We would expect a longer budget to close it eventually; what
we can report is that it does not close within the $300$\,s cap, which is eight
to forty-four times our own runtime depending on the instance.

Two caveats keep this in proportion. First, the effect depends on
$\varphi > 0$: Section~\ref{sec:phizero} repeats the whole comparison with
dispatch free, and there the two are indistinguishable at matched wall-clock, so
ALNS would be expected to pass us given more time. Second, the
comparison uses the standard operator set of~\citet{ropke2006adaptive}; an
implementation tuned for these sizes, or one carrying a route-elimination
operator, would be expected to do better than this.

\subsection{The parallel assignment}\label{sec:lagrangianresults}

Section~\ref{sec:lagrangian} replaces the one coupled program by one
subproblem per courier. \tabref{tab:lagrangian} reports what that costs, in
components rather than a single percentage, on the same 18 instances as
\tabref{tab:main}.

\begin{table}[t]\centering
\sbox{\tabbox}{\small\begin{tabular}{@{}r@{\;\;} l@{\;} r@{\;\;}r@{\;\;}r@{\;\;} r@{\;\;} r@{\;\;} r@{}}\toprule
$|\mathcal{R}|$ & assignment & travel & tard. & cour. & cost & assign. & $\Delta$\\\midrule
$40$ & exact program & 624 & 18 & 8.0 & 1441 & 0.2\,s & ---\\
 & multipliers only & 547 & 6 & 10.0 & 1553 & 0.1\,s & $+7.8\%$\\
 & $+$ fleet refit & 593 & 13 & 8.7 & 1472 & 0.1\,s & $+2.2\%$\\
 & $+$ shortlist refit & 614 & 10 & 8.3 & 1457 & 0.3\,s & $+1.1\%$\\
\addlinespace[2pt]
$100$ & exact program & 1121 & 67 & 20.7 & 3255 & 3.3\,s & ---\\
 & multipliers only & 1078 & 11 & 24.5 & 3539 & 0.3\,s & $+8.7\%$\\
 & $+$ fleet refit & 1057 & 17 & 23.7 & 3441 & 0.3\,s & $+5.7\%$\\
 & $+$ shortlist refit & 1096 & 81 & 21.2 & 3294 & 1.5\,s & $+1.2\%$\\
\addlinespace[2pt]
$200$ & exact program & 1874 & 231 & 40.8 & 6189 & 10.5\,s & ---\\
 & multipliers only & 1830 & 34 & 46.5 & 6515 & 0.5\,s & $+5.3\%$\\
 & $+$ fleet refit & 1850 & 72 & 46.0 & 6522 & 0.6\,s & $+5.5\%$\\
 & $+$ shortlist refit & 1870 & 204 & 41.2 & 6191 & 6.6\,s & $+0.1\%$\\
\addlinespace[2pt]
\midrule
\emph{all} & exact program & 1206 & 105 & 23.2 & 3628 & 4.7\,s & ---\\
 & multipliers only & 1152 & 17 & 27.0 & 3869 & 0.3\,s & $+7.3\%$\\
 & $+$ fleet refit & 1166 & 34 & 26.1 & 3812 & 0.3\,s & $+4.5\%$\\
 & $+$ shortlist refit & 1193 & 98 & 23.6 & 3647 & 2.8\,s & $+0.8\%$\\
\bottomrule\end{tabular}}%
\setlength{\tabw}{0.80\textwidth}%
\ifdim\wd\tabbox>\tabw\setlength{\tabw}{\wd\tabbox}\fi
\begin{minipage}{\tabw}
\centering
\caption{$|\mathcal{R}| = 40, 100, 200$ requests, $|\mathcal{K}| = 12, 30, 60$ couriers, $Q_k = 5$, $\varphi = 100$, $\rho = 1$. The cost of distributing the assignment, on the instances of \tabref{tab:main} at $\varphi = 100$. \emph{Assign.} is the time in the assignment step, the only step that couples couriers; $\Delta$ is the paired cost difference against the exact program. Each block is one problem size; \emph{all} pools the instances of every size, which is what the text quotes.}
\label{tab:lagrangian}
\small\begin{tabular*}{\tabw}{@{\extracolsep{\fill}}r@{\;\;} l@{\;} r@{\;\;}r@{\;\;}r@{\;\;} r@{\;\;} r@{\;\;} r@{}}\toprule
$|\mathcal{R}|$ & assignment & travel & tard. & cour. & cost & assign. & $\Delta$\\\midrule
$40$ & exact program & 624 & 18 & 8.0 & 1441 & 0.2\,s & ---\\
 & multipliers only & 547 & 6 & 10.0 & 1553 & 0.1\,s & $+7.8\%$\\
 & $+$ fleet refit & 593 & 13 & 8.7 & 1472 & 0.1\,s & $+2.2\%$\\
 & $+$ shortlist refit & 614 & 10 & 8.3 & 1457 & 0.3\,s & $+1.1\%$\\
\addlinespace[2pt]
$100$ & exact program & 1121 & 67 & 20.7 & 3255 & 3.3\,s & ---\\
 & multipliers only & 1078 & 11 & 24.5 & 3539 & 0.3\,s & $+8.7\%$\\
 & $+$ fleet refit & 1057 & 17 & 23.7 & 3441 & 0.3\,s & $+5.7\%$\\
 & $+$ shortlist refit & 1096 & 81 & 21.2 & 3294 & 1.5\,s & $+1.2\%$\\
\addlinespace[2pt]
$200$ & exact program & 1874 & 231 & 40.8 & 6189 & 10.5\,s & ---\\
 & multipliers only & 1830 & 34 & 46.5 & 6515 & 0.5\,s & $+5.3\%$\\
 & $+$ fleet refit & 1850 & 72 & 46.0 & 6522 & 0.6\,s & $+5.5\%$\\
 & $+$ shortlist refit & 1870 & 204 & 41.2 & 6191 & 6.6\,s & $+0.1\%$\\
\addlinespace[2pt]
\midrule
\emph{all} & exact program & 1206 & 105 & 23.2 & 3628 & 4.7\,s & ---\\
 & multipliers only & 1152 & 17 & 27.0 & 3869 & 0.3\,s & $+7.3\%$\\
 & $+$ fleet refit & 1166 & 34 & 26.1 & 3812 & 0.3\,s & $+4.5\%$\\
 & $+$ shortlist refit & 1193 & 98 & 23.6 & 3647 & 2.8\,s & $+0.8\%$\\
\bottomrule\end{tabular*}
\end{minipage}
\end{table}

The multipliers $\lambda$ of~\eqref{eq:lagr} coordinate one of the two
decisions well and the other badly: which courier takes a request, and how many
couriers open. The components separate
them. The Lagrangian variant travels \emph{less} than the exact program
(1152 against 1206) and is \emph{more}
punctual (17 against 105); it loses
only on the fleet, opening 27.0 couriers against
23.2, and that one term turns a better route plan into a
$+7.3\%$ worse solution (0/18 wins,
$p<10^{-3}$). This is the expected behaviour of a Lagrangian relaxation of a
fixed-charge problem: a multiplier on the covering constraint tells a courier
which requests are cheap for it, and nothing tells the couriers collectively how
many of them should open.

Two repairs take the multipliers as a starting point rather than as the answer,
and both appear in \tabref{tab:lagrangian}. The \emph{fleet refit} keeps the
set of couriers the multipliers opened, together with the two sizes either side
of it, and assigns requests to each of those fleets in turn; with the open set
fixed the dispatch charge is a constant and what remains is a transportation
problem, so each trial is a linear program. The \emph{shortlist refit} leaves
the fleet decision open and uses the reduced costs only to narrow the problem,
keeping for each request the three couriers of lowest reduced cost and solving
the exact assignment program on what is left.

The second recovers most of the loss, $+0.8\%$ pooled, and more of
it the larger the instance:
$+1.1\%$ at $|\mathcal{R}|=40$ against
$+0.1\%$ at $|\mathcal{R}|=200$. That confirms which
term carries the gap rather than removing it. That repair is not free: the
program it solves is smaller than the original but still an integer program, and
still coupled, which costs
2.77\,s against the exact program's
4.66\,s pooled, but only because it is cheaper at the two
larger sizes; at $|\mathcal{R}|=40$, where the exact program
closes in 0.17\,s, the repair is the dearer of the two at
0.26\,s. What the distributed form buys is the coupled step:
16 times less serial time
pooled, and a saving that grows with the instance, from
1.5 times at $|\mathcal{R}|=40$ (0.17\,s against 0.11\,s) to
21 times at $|\mathcal{R}|=200$, in a step that
needs no information about other couriers. We therefore use the
exact assignment in the proposed method, and report the Lagrangian variant as
the option for a setting where the assignment must itself be distributed.

% ===== END GENERATED RESULTS =====

\section{Conclusion}\label{sec:conclusion}
A cluster-first decomposition is only as good as the quantity its grouping
optimises. We have argued, and measured, that the natural choice, proximity in
space and time, is wrong in its temporal half: a courier serves its group
sequentially, so requests that fall due together are in contention rather than
in affinity. We replace that criterion with the marginal cost of the induced
routing subproblem, and we let the one integer program that groups the requests
also decide which couriers to open.

Against the best of six published clustering rules, taken instance by instance,
our proposed method is $29\%$ cheaper with those rules as published and $12\%$
cheaper once they carry our own improvement step, on every one of the 18
instances in both comparisons, and the margin widens as the fleet grows. It does
not rest on the dispatch charge, and with couriers free it is wider still.
Because the grouping confines sequencing to a few requests per courier, we keep
the expensive work decomposable by courier, a structure we measure but
do not exploit, since every runtime we report is single-core. With
candidate lists we solve a
thousand requests and three hundred couriers in about a minute, sixteen times
faster than the strongest clustering alternative and $34\%$ cheaper. We return
the same answer whatever the random seed.

We bound the claim in three ways. Our margin is smaller on benchmark batches we
did not design, and Section~\ref{sec:external} traces the reason to instance
structure rather than to size, since the gain is largest where a request's pickup and delivery lie close together, which
is the regime courier dispatch operates in and not the one those batches
represent. Against a large-neighbourhood search we are ahead at matched
wall-clock when dispatch is charged and level when it is free, and we do not see
it reach our cost within eight to forty-four times our budget, which says what one
improvement costs a search over complete solutions rather than where either
method's quality ceiling lies. Distributing the assignment through multipliers
costs $7.3\%$, all of it in the fleet decision.

\section*{Declaration of generative AI and AI-assisted technologies in the
manuscript preparation process}

During the preparation of this work the authors used Claude (Anthropic) to
generate the tables and figures from the simulation results that the authors
coded and ran, to improve the phrasing of the manuscript for clarity and
readability, and to screen the manuscript for typographical and grammatical
errors. After using this tool, the authors reviewed and edited the content as
needed and take full responsibility for the content of the published article.

\bibliographystyle{elsarticle-harv}
\bibliography{bibFiles/references}

\end{document}